\documentclass[a4paper,11pt]{article}
\usepackage{jheppub} 
\usepackage{lineno}
\usepackage{float}
\newcommand{\tr}{\operatorname{tr}}
\newcommand{\tl}{\tilde{\ell}}
\usepackage{inconsolata}

\title{\boldmath Entanglement entropy in type II$_1$ von Neumann algebra: examples in Double-Scaled SYK}

\author[a]{Haifeng Tang}
\affiliation[a]{Department of Physics, Stanford University, Stanford, California 94305, USA}
\emailAdd{hftang@stanford.edu}

\abstract{
An intriguing feature of type II$_1$ von Neumann algebra is that the entropy of the mixed states is negative. Although the type classification of von Neumann algebra and its consequence in holography have been extensively explored recently, there has not been an explicit calculation of entropy in some physically interesting models with type II$_1$ algebra. In this paper, we study the entanglement entropy $S_n$ of the fixed length state $\{|n\rangle\}$ in Double-Scaled Sachdev-Ye-Kitaev model, which has been recently shown to exhibit type II$_1$ von Neumann algebra. These states furnish an orthogonal basis for 0-particle chord Hilbert space. We systematically study $S_n$ and its R\'enyi generalizations $S_n^{(m)}$ in various limit of DSSYK model, ranging $q\in[0,1]$. We obtain exotic analytical expressions for the scaling behavior of $S_n^{(m)}$ at large $n$ for random matrix theory limit ($q=0$) and SYK$_2$ limit ($q=1$), for the former we observe highly non-flat entanglement spectrum. We then dive into triple scaling limits where the fixed chord number states become the geodesic wormholes with definite length connecting left/right AdS$_2$ boundary in Jackiw-Teitelboim gravity. In semi-classical regime, we match the boundary calculation of entanglement entropy with the dilaton value at the center of geodesic, as a nontrivial check of the Ryu-Takayanagi formula.
}


\begin{document}
\maketitle
\flushbottom

\section{Introduction}
\paragraph{Motivations.}

von Neumann algebra has recently received much attention from the high energy theory community due to its improvement of our understanding of algebras underlying different scenarios of bulk emergence in holography~\cite{Witten:2018zxz,Witten:2021jzq,Witten:2021unn,Chandrasekaran:2022cip,Chandrasekaran:2022eqq,Penington:2023dql,Witten:2023qsv,Strohmaier:2023opz,Witten:2023xze,Jensen:2023yxy,Sorce:2023fdx,Xu:2024hoc}. The classification of von Neumann algebra into three distinct types has deep relation with the algebra of observer exterior to a horizon, possibly from eternal black hole or cosmic inflation.

The type I von Neumann algebra is what we are most familiar with, which, roughly speaking, describes the matrix algebra of quantum mechanics with (in)finite dimensional Hilbert. Mathematically speaking, there is well-defined trace and both pure and mixed states, and of course the notion of entanglement entropy. The next familiar case is the type III algebra, one of whose sub-classification, type III$_1$, describes the algebra of local operators in QFT defined on fixed spacetime metric background. There is no useful definition of trace therefore no rigorous definition of entropy there.

The most unexpected category is type II von Neumann algebra, which has no pure states but all mixed states, and the definition of trace exists. Type II$_{\infty}$ describes the algebra of observer in the exterior region of a quantum black hole~\cite{Chandrasekaran:2022eqq} and type II$_1$ is believed to describe the algebra of observer in the static patch of de Sitter space~\cite{Chandrasekaran:2022cip}.

One particularly interesting feature of type II$_1$ algebra is that the entropy of density matrices is all negative. Physically this is because every density matrix in this infinite dimensional Hilbert space is close to the maximally mixed state, therefore its entropy (or the entanglement entropy between its purifier system) diverges. However, its relative value compared to the entropy of maximal mixed state remains finite. So it is reasonable to redefine this finite deficit value as entropy. Mathematically speaking, this is because the definition of `type II trace' differs from the usual `type I trace' by a diverging multiplicative constant, which in turn leads to the positive diverging additive constant to entropy of all states. Since this divergent is universal, we may simply drop it in the same spirit with usual thermodynamics where the thermal entropy is well defined up to an additive constant. See~\cite{Witten:2018zxz,Witten:2021jzq} for more physical intuitions for classification of von Neumann algebras and see~\cite{Sorce:2023fdx} for a pedagogical and self-contained introduction to those more mathematically rigorous aspects.

Despite the intriguing features of entropy in type II$_1$ von Neumann algebra, to our knowledge, it has not been calculated explicitly in some models that realize such an algebra. In this work, we provide an explicit calculation of entropy of physically relevant states in DSSYK model.

Double Scaled Sachdev-Ye-Kitaev (DSSYK) model~\cite{Berkooz2018,Berkooz2019,Berkooz2020SUSY,Berkooz2021complex,Berkooz2022Quantum,Mukhametzhanov2023tcg,Okuyama2023bch,Okuyama2023iwu,Okuyama2023byh,Goel2023svz,Boruch2023,Verlinde:2024zrh,Verlinde:2024znh,Narovlansky:2023lf,Milekhin:2023bjv} has been conjectured previously~\cite{Lin2022} and rigorously shown very recently~\cite{Xu:2024hoc} to support a type II$_1$ algebra. This model has also been recently conjectured to be holographically dual to observers traveling in the static patch of de Sitter space, possibly on the stretched horizon~\cite{Susskind2021omt,Susskind2021dfc,Susskind2021esx,Susskind2022fop,Susskind2022dfz,SusskindLin2022nss,Susskind2022bia,Susskind2023hnj,Susskind2023rxm} or at the podes~\cite{Verlinde:2024zrh,Verlinde:2024znh,Narovlansky:2023lf}.

\paragraph{Model considered.}
In this work, we provide an explicit calculation of the entanglement entropy $S_n$ (in two-sided picture, or the normal entropy of mixed states in one-sided picture) of fixed-chord-number states $\{|n\rangle,n\geq0\}$. In the construction of Hilbert space of DSSYK, they furnish an orthogonal basis for 0-particle subspace. In the triple scaling limits, this set of states describes the fixed-length-states $\{|\tl\rangle\}$ in Jackiw-Teitelboim gravity~\cite{Jackiw1984je,Teitelboim1983ux,Polchinski2015,jensen2016,yang2016,Mertens2022,Saad:2019lba,
Jafferis:2019wkd,Harlow:2018tqv}, whose wave-function in energy basis $\langle E|\tl\rangle$ is obtained by path integral of JT gravity on a disk configuration where boundary condition on the asymptotic boundary is fixed energy $E$ and the boundary condition in the bulk line is a geodesic with renormalized length $\tl$. Physically speaking, $|\tl\rangle$ describes the wave function of wormhole (in the sense of ER$=$EPR) connecting left and right boundaries of AdS$_2$.  We compare the calculation from microscopic DSSYK model in triple scaling limit with semi-classical calculation in the bulk of 2D JT gravity and find that the entropy matches the on-shell value of dilaton field at the center of the geodesic, which is a manifestation of Ryu-Takayanagi formula~\cite{Takayanagi2006}.

\paragraph{Summary of the main results.}

We calculate $S_n$ and its $m^{\text{th}}$-R\'enyi generalization $S^{(m)}_n$(for $m=1$, R\'enyi entropy goes back to entanglement entropy, or the von Neumann entropy $S_n$) as a function of $n$ in various limits of $q$-parameter in DSSYK model. We would see that $S_n^{(m)}$ decreases from zero as $n$ increases, furnishing an explicit manifestation of entropy in type II$_1$ algebra.

In section~\ref{sec:set up}, we briefly introduce DSSYK model and the formulation of 0-particle Hilbert space.

In section~\ref{sec:q=0}, we calculate $S_n^{(m)}$ in $q=0$ limit of DSSYK, which we call RMT (Random Matrix Theory) limit, since the density of states of DSSYK there approaches Wigner semicircle. We find that the large $n$ scaling behaviour is quite different for different R\'enyi index $m$:
\begin{equation}
q=0,\ S_n^{(m)}\left\{
\begin{aligned}
&=-1+\frac{1}{n+1},\ &m=1\\
&\approx -a_m+b_m(n+1)^{-(3-2m)},\ & 1<m<\frac{3}{2}\\
&\approx-2\log\log(n+1),\ &m=\frac{3}{2}\\
&\approx-c_m\log(n+1),\ &m>\frac{3}{2}
\end{aligned}
\right.
\end{equation}
where $a_m,b_m,c_m$ are some positive coefficients depend on $m$ but not on $n$. For $m=1$ the result is exact for all $n\geq0,n\in\mathbb Z$, while for other ranges of $m$ the above `$\approx$' means asymptotic behaviour of large $n$. A byproduct is that we also know the exact expression at $m=2$: $S_n^{(m=2)}=-\log(n+1)$, which is consistent with the scaling behaviour. 

Such exotic scaling behaviours would indicate a highly non-flat entanglement spectrum. We see that for $1\leq m<\frac{3}{2}$, $S_n^{(m)}$ remains finite when $n$ goes to infinity, while for $m>\frac{3}{2}$, $S_n^{(m)}$ diverges when $n$ goes to infinity. 

From a quantum information perspective, the qualitatively different behaviour of von Neumann entropy and R\'enyi entropy means the latter is not always a good entanglement measure. By contrast, in the large dimension limit of many random tensor network models, the entanglement spectrum is flat. Hence, people often use second R\'enyi entropy ($m=2$), which is easier to handle, to represent the behaviour of entanglement entropy ($m=1$).

In section~\ref{sec:q=1}, we calculate $S_n^{(m)}$ in $q=1$ limit (while keeping $n$ finite), which we dub as SYK$_2$ limit, since the density-of-state there approaches Gaussian distribution. We find that at large $n$, $S_n^{(m)}$ linearly decrease with $n$ for all $m\geq1$:
\begin{equation}
q=1,\ S_n^{(m)}\approx-d_m\cdot n
\end{equation}
where $d_m$ is some positive coefficient depend on $m$. This indicates a less interesting entanglement spectrum.

In section~\ref{sec:triple scaling}, we focus on the entanglement entropy $S_n\equiv S_n^{(m=1)}$ of DSSYK model in triple scaling limit: $q\rightarrow1^{-},n\rightarrow+\infty$, while $\tl$ remains finite. Parametrize $q\equiv e^{-\lambda},\lambda\in[0,+\infty]$, the remormalized length $\tl$ is defined by $e^{-\lambda n}=\lambda^2e^{-\tl}$. This is equivalent to $\lambda\rightarrow0+$. To do this, we need some preparation first. 

In section~\ref{sec:intermediate range}, we first show the numerical result of $S_n$ when changing $q$ from $0$ to $1$. In section~\ref{sec:plateau value}, we calculate another limit: $\lambda$ is kept small but finite, and then taking $n\rightarrow+\infty$. The entropy is given by:
\begin{equation}
S_{n=\infty}=-\frac{\pi^2}{6}\lambda^{-1}+\frac{1}{2}\log(\lambda^{-1})+\frac{1}{2}\log(2\pi)-1-\frac{1}{12}\lambda+\frac{\lambda}{2\pi^2}\operatorname{Li}_2(e^{-4\pi\lambda^{-1}})
\label{eq:1.3}
\end{equation}

In section~\ref{sec:estimation of triple scaling}, we estimate the entropy in the triple scaling limit. We find that $\Delta S(\tl)=S(\tl)-S_{n=\infty}$ is finite and positive when taking $\lambda\rightarrow0^+$ (by contrast, $S_{n=\infty}$ itself diverges to minus infinity according to equation~(\ref{eq:1.3})). Upon reasonable estimation, we argue that $\Delta S(\tl)$ is positively related to a particular energy scale, the `\textit{penetration energy}' $k_0(\tl)$, of the low energy wave function which is described by Liouville quantum mechanics. Since $k_0(\tl)$ decreases with $\tl$, this means the `area' of the wormhole (this notion of `area' is from the dimension reduction of higher dimensional theory~\cite{Mertens2022}. In pure 2D JT gravity, the entropy is represented by dilation field) decreases with its length, which is reasonable from the bulk point of view. 

In section~\ref{sec:matching JT} we compare the calculation from the microscopic DSSYK model in triple scaling limit with semi-classical calculation in the bulk of 2D JT gravity and find that the entropy $\Delta S(\tl)$ matches the on-shell value of dilaton field at the center of the geodesic.

In section~\ref{label:conclusion} we summarize our result and discuss some open questions.

\section{Setup of entropy calculation}
\label{sec:set up}
\paragraph{Brief introduction to DSSYK model.} For readers' convenience, we first very briefly review the definition of DSSYK model and remark on different parameter regions that are relevant to our calculation.

Consider a (0+1) dimensional quantum mechanical model consisting of $N$ Majorana fermions with the following commutation relation:
\begin{equation}
\{\chi_i,\chi_j \}=2\delta_{i,j},\ i,j=1,...,N
\end{equation}
The Hamiltonian of system contains $p$-local all-to-all connected interaction between Majorana with random coupling coefficients~\cite{kitaev2015,sachdev1993,maldacena2016remarks,Polchinski2016,altland2016,altland2017}:
\begin{equation}
H=i^{p/2}\sum_{i_1<...<i_p}J_{i_1...i_p}\chi_{i_1}\cdots\chi_{i_p}
\end{equation}
where $J_{i_1...i_p}\equiv J_{I}$ is are independent identically distributed Gaussian variable with zero mean. Their variance is given by~\cite{Berkooz2018}:
\begin{equation}
\langle J_I\rangle=0,\ \langle J_IJ_J\rangle=(C_N^p)^{-1}\delta_{IJ}
\end{equation}
where $I\equiv (i_1,...,i_p)$ is a collective index for notational simplicity. Next, we define the parameter $q,\lambda$ which charaterize the localness of interaction:
\begin{equation}
\lambda\equiv\frac{2p^2}{N},\ q\equiv e^{-\lambda}
\end{equation}

DSSYK model is defined by taking the following limit of parameter: $N\rightarrow+\infty,p\rightarrow+\infty,\lambda\rightarrow\text{Const}$. In other word, $p\sim O(\sqrt{N})$. 

It turns out that many calculations can be simplified in this region and obtain many interesting analytical results, including partition function, two-point and four-point matter correlation functions~\cite{Berkooz2018,Berkooz2019}. Their basic ingredient, which is also relevant to our work, is the calculation of moments of Hamiltonian: $\mu_{2n}\equiv\tr(H^{2n})$. They can be elegantly evaluated using chord diagram techniques~\cite{Berkooz2018,Berkooz2019}.

By tuning parameter $q$ in the range of $q\in[0,1]$, DSSYK model can be related to many other models, smoothly interpolating between random matrix theory and the usual large-$p$ SYK model and SYK$_2$ model:
\begin{itemize}
\item[1.]\textit{Random Matrix Theory limit.} When $q=0$, the density-of-state approaches Wigner semi-circle, which is the large $N$ limit of Gaussian Unitary Ensemble (GUE)~\cite{mehta2004random}. This is reasonable because when $q=0$, we are making the interaction range completely non-local by taking $\lambda=+\infty$, which should resemble a random matrix in certain ensembles.
\item[2.]\textit{SYK$_2$ limit.} This is achieved by taking $q=1$ while keeping $n$ finite. In this limit, we are effectively focusing on the center of spectrum, which is a Gaussian distribution. Here, $\mu_{2n}$ matches the moment of SYK$_2$ model in large $N$ limit~\cite{Garcia-Garcia:2018fns}.
\item[3.]\textit{Triple scaling limit.} This is obtained by taking $q\rightarrow1^-,n\rightarrow+\infty$ while keeping $\tl$ finite, where the renormalized length $\tl$ is defined by $e^{-\lambda n}=\lambda^2e^{-\tl}$. The low energy behavior of DSSYK in triple scaling limit is described by Liouville quantum mechanics, the same as the boundary dynamics of JT gravity via covariant quantization~\cite{Harlow:2018tqv}.
\end{itemize}

\paragraph{Review of construction of 0-particle Hilbert space.} In recent papers~\cite{Lin2022,Lin2023}, the authors had constructed explicit boundary states supported on two-sided Hilbert space whose bulk dual is interpreted as a wormhole state with definite length. In this work, by assuming the applicability of Ryu-Takayanagi formula~\cite{Takayanagi2006}, we provide a boundary characterization of wormhole's area (minimal area of the throat) through entanglement entropy between left-side and right-side of those wormhole states.

\begin{figure}[t]
    \centering
    \includegraphics[width=\textwidth]{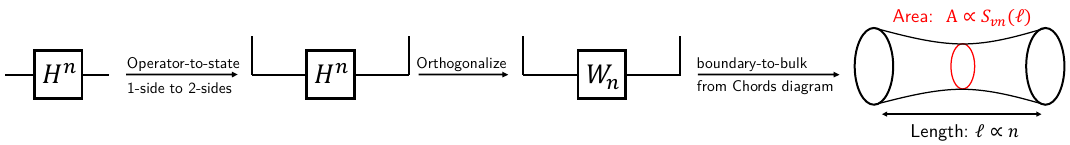}
    \caption{A schematic summary of the bulk-to-boundary holographic map constructed in~\cite{Lin2022}. In this paper, we are interested in the `area'(dilaton value in 2D gravity) of the wormhole states, which is characterized by the entanglement entropy in boundary's perspective. }
    \label{fig:schematic plot}
\end{figure}

In the bulk reconstruction process of~\cite{Lin2022}, the operator $H^n$, can be interpreted as a pure state $|H^n\rangle$ on doubled Hilbert space (Hilbert space of the wormhole) $\mathcal{H}\otimes\mathcal{H}^*$. Roughly speaking, this state $|H^n\rangle$ describes a wormhole with length $\tl\propto n$. However, to obtain the wormhole state with \textit{definite} length, we need to further orthogonalize the set of states $\{|H^n\rangle\}\rightarrow\{|W_n\rangle\}$. Then, finally, we interpret $|W_n\rangle$ to be the state with definite length $\tl\propto n$. This procedure is schematically summarized in figure~\ref{fig:schematic plot}.

The orthogalisation procedure is the Lanczos algorithm:
\begin{align}
&|W_0\rangle=|\mathbb{I}\rangle,\\
&|W_1\rangle=\frac{1}{b_1}|HW_0\rangle,\\
&|W_n\rangle=\frac{1}{b_n}\bigg(|HW_{n-1}\rangle-b_{n-1}|W_{n-2}\rangle \bigg),\ \ \text{for }n\geq2
\end{align}
where $b_n$ is determined step-by-step from requiring normalization condition $\langle W_n|W_n\rangle=1$, and the inner product is defined as $\langle A|B\rangle=\tr(A^{\dagger}B)$. Here following the convention in DSSYK literature, we normalize $\tr\mathbb{I}=1$, where $\mathbb{I}$ is the identity operator.

One can recursively show that upon the above construction, $\{|W_n\rangle\}$ becomes orthonormal~\cite{Parker2019}: $\langle W_m|W_n\rangle=\delta_{mn}$. We also notice that $W_n$ is essentially a real polynomial of $H$ with the highest power $n$. Then $\{b_n\}$ determined from $\tr(W_n^2)=1$ only require knowledge of the moments of Hamiltonian, namely $\{\mu_{2n}\}=\{\tr(H^{2n})\}$. The latter is well-known calculated by Berkooz using chords diagram~\cite{Berkooz2018}. For example, we can explicitly write down the first three coefficients:
\begin{align}
&W_1=\frac{1}{b_1}H && \longrightarrow \tr H^2=\mu_{2}=b_1^2\\
&W_2=\frac{1}{b_1b_2}(H^2-b_1^2\mathbb{I}) && \longrightarrow \tr H^4=\mu_{4}=b_1^4+b_1^2b_2^2\\
&W_3=\frac{1}{b_1b_2b_3}(H^3-(b_1^2+b_2^2)H) && \longrightarrow \tr H^6=\mu_6=b_1^6+2b_1^4b_2^2+b_1^2b_2^4+b_1^2b_2^2b_3^2
\end{align}
The relation between $\{\mu_{2n}\}$ and $\{b_n\}$ is exactly the hopping problem on the Krylov chain in the context of study on Krylov complexity~\cite{Parker2019,Bhattacharyya2023,Erdmenger2023,Kar2022,Rabinovici2021,Jian2021,Rabinovici20221,Rabinovici20222,Liu2023,Tang:2023ocr}, and one can check that the number of terms on RHS ($2b_1^4b_2^2$ is counted as two terms) of $\mu_{2n}$ is $C_n=\frac{(2n)!}{(n+1)!n!}$, with $C_n$ the Catalan number, which counts the number of Dicke path. We also notice that in the chords diagram result~\cite{Berkooz2018}, $\mu_{2n}=\langle0|\tilde{T}^{2n}|0\rangle$ has already been tri-diagonalized where the symmetric transfer matrix read as~\cite{Berkooz2018}:
\begin{equation}
\tilde{T}=
\begin{bmatrix}
0 & \sqrt{\frac{1-q}{1-q}} & 0 & 0 &  \cdots\\
\sqrt{\frac{1-q}{1-q}} & 0 & \sqrt{\frac{1-q^2}{1-q}} & 0 &  \cdots\\
0 & \sqrt{\frac{1-q^2}{1-q}} & 0 & \sqrt{\frac{1-q^3}{1-q}} & \cdots \\
0 & 0 & \sqrt{\frac{1-q^3}{1-q}} & 0 & \cdots\\
\vdots & \vdots & \vdots & \vdots & \ddots
\end{bmatrix}
\label{eq:tranfer matrix}
\end{equation}
Then we can read out $b_n$ directly from the hopping coefficient on the link:
\begin{equation}
b_n=\sqrt{\frac{1-q^n}{1-q}}
\end{equation}
Next, we work out the explicit form of $W_n$ as a polynomial of $H$. Since for later convenience, the entropy calculation only depends on the spectrum of $H$, so we can safely replace $H$ with its eigenvalue~\cite{Berkooz2018} $H\rightarrow E=\frac{2\cos\theta}{\sqrt{1-q}}$. From the recursion relation $b_nW_n+b_{n-1}W_{n-2}=HW_{n-1}$, we can define a new variable:
\begin{equation}
u_n\equiv(1-q)^{n/2}b_nb_{n-1}\cdots b_1W_n\longrightarrow (1-q^{n})u_{n-1}+u_{n+1}=2\cos\theta u_n
\end{equation}
The solution of this recursion relation with initial condition $u_0=1$ is just the $q$-Hermite polynomial $H_n(z|q)$:
\begin{equation}
u_n=H_n(\cos\theta|q)=\sum_{k=0}^{n} \frac{(q;q)_n}{(q;q)_{k}(q;q)_{n-k}}e^{i(n-2k)\theta}
\end{equation}
where $(a;q)_n\equiv(1-aq^0)(1-aq^1)\cdots(1-aq^{n-1})$ is $q$-Pochhammer symbol. Here, $H_n(z|q)$ is degree $n$ real polynomial of $z$. Then $W_n$ as a degree $n$ polynomial of $H$ is given by:
\begin{equation}
W_n(H)=\frac{H_n(\sqrt{1-q}H/2|q)}{\sqrt{(q;q)_n}}
\end{equation}

\paragraph{Set up for calculation of entropy.}Then, we are ready to calculate the reduced density matrix $\rho_n$ on left side:
\begin{equation}
\rho_n=\tr_{R}|W_n\rangle\langle W_n|=W_n^2
\end{equation}
Then the von Neumann entropy reads as:
\begin{equation}
S_{n}=-\tr \rho_n\log\rho_n=-\int_{0}^{\pi}d\theta \cdot\Psi(\theta,q) \frac{H_n(\cos\theta|q)^2}{(q;q)_n}\log\left[\frac{H_n(\cos\theta|q)^2}{(q;q)_n}\right]
\label{eq:mutual information}
\end{equation}
where the eigenvalue is parametrized by $E(\theta)=\frac{2\cos\theta}{\sqrt{1-q}}$ and the distribution function $\Psi(\theta,q)$ is given by~\cite{Berkooz2018}:
\begin{equation}
\Psi(\theta,q)=\frac{1}{2\pi}(q;q)_{\infty}(e^{+2i\theta};q)_{\infty}(e^{-2i\theta};q)_{\infty}
\label{eq:density if state}
\end{equation}

\paragraph{Natural emergence of negative entropy characterizing type II$_1$ algebra.}

To show that the entropy is negative, an interesting observation is that we can express equation~(\ref{eq:mutual information}) in terms of relative entropy (Kullback-Leibler Divergence) of some classical distribution. We notice that the $q$-Hermite polynomial can be expressed in terms of orthonormal wave function:
\begin{equation}
\begin{aligned}
&\psi_n(\cos\theta|q)=\psi_0(\cos\theta|q)\frac{H_n(\cos\theta|q)}{\sqrt{(q;q)_{n}}}\\
&\psi_0(\cos\theta|q)=\frac{\sqrt{(q;q)_{\infty}}|(e^{2i\theta};q)_{\infty}|}{\sqrt{2\pi}}=\sqrt{\Psi(\theta,q)}\\
&\delta_{mn}=\int_0^{\pi}d\theta\ \psi_n(\cos\theta|q)\psi_m(\cos\theta|q)
\end{aligned}
\end{equation}
Then we have $\rho_n=\frac{\psi_n^2}{\psi_0^2}$. As a result:
\begin{equation}
S_n=-\int_{0}^{\pi}d\theta\ \psi_n(\cos\theta|q)^2\log\left[\frac{\psi_n(\cos\theta|q)^2}{\psi_0(\cos\theta|q)^2}\right]=-D_{KL}(\psi_n^2||\psi_0^2)
\label{eq:D_KL}
\end{equation}
We notice that $\psi_n^2$ are well-defined classical probability distributions on $\theta\in[0,\pi]$. Since KL-divergence is always positive, this indicates the entropy is negative, indicating the type II$_1$ nature of algebra.

The reason for negative entropy lies in the fact that we take the normalization condition $\tr\mathbb{I}=\tr(H^0)=1$ for notational simplicity in DSSYK literature. This means we are secretly using a `type II$_1$ trace'. To re-obtain a physical entropy, we need to compensate a divergent multiplicative factor $\sqrt{2}^N$ to obtain the usual `type I trace'. This step brings back the universal divergent additive constant to entropies: $S_n(\text{type I})=S_n(\text{type II}_1)+N\log\sqrt{2}$. We notice that since we are working at $N\rightarrow\infty$ limit ($N$ is the number of Majorana fermion in microscopic DSSYK model), so $S_n(\text{type I})=+\infty$. Therefore, it is meaningful to study its relative value to the maximal value, namely $S_n(\text{type II}_1)$, which is finite in $N\rightarrow\infty$ limit. This exactly brings us back to the type II$_1$ entropy.

Back to our main theme, for later convenience, we also study R\'enyi version of entanglement entropy defined by:
\begin{equation}
S^{(m)}_n=\frac{1}{1-m}\log[\tr\rho_n^m]=\frac{1}{1-m}\log\left(\int_{0}^{\pi}d\theta \Psi(\theta,q)\left[\frac{H_n(\cos\theta|q)^2}{(q;q)_n}\right]^{m}\right)
\end{equation}
where, by taking $m\rightarrow1$ we recover von Neumann entanglement entropy defined above. We shall see that the entanglement spectrum has highly non-trivial dependence on R\'enyi index $m$.

\section{RMT limit: $q=0$}
\label{sec:q=0}
We first consider the RMT (Random Matrix Theory) when we take $q\rightarrow0$. This is obviously seen from the distribution function $\Psi(\theta,q)$, approaching the Wigner semicircle law~\cite{mehta2004random} in this limit. Alternatively, for finite $q$, we also notice that the RMT behavior shows up when $n\rightarrow+\infty$ where we can safely replace $q^n\approx0$.
\paragraph{Numerical results and naive calculation }For concreteness, we take $q=0$. The $q$-Pochhammer symbol simplified to be $(q;q)_{n}=1$ therefore $q$-Hermite polynomial and distribution function are simplified as:
\begin{align}
H_n(\cos\theta|q)&=\sum_{k=0}^{n} e^{i(n-2k)\theta}=\frac{\sin (n+1)\theta}{\sin\theta}\\
\Psi(\theta,q)&=\frac{1}{2\pi}(1-e^{2i\theta})(1-e^{-2i\theta})=\frac{2}{\pi}\sin^2\theta
\end{align}
Then the R\'enyi entropy is given by:
\begin{equation}
S_n^{(m)}=\frac{1}{1-m}\log\left( \int_{0}^{\pi}d\theta\frac{2}{\pi}\sin^2\theta\left[\frac{\sin (n+1)\theta}{\sin\theta}\right]^{2m}\right)
\label{eq:1}
\end{equation}

\begin{figure}[t] 
    \centering
    \includegraphics[width=\textwidth]{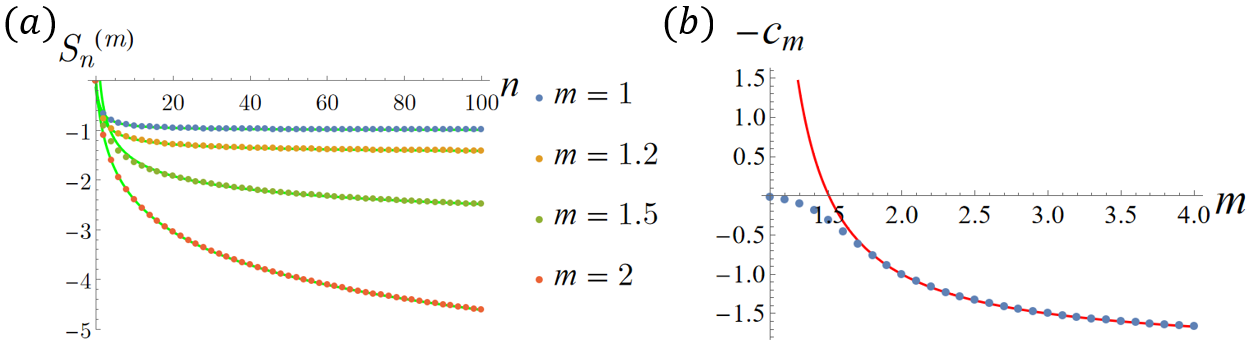}
    \caption{(a) Entanglement entropy and R\'enyi entropy at RMT limit ($q=0$). The colored dots are $S_{n}^{(m)}$ from equation~(\ref{eq:1}). (b) We linearly fit the curve with ansatz form $a_m-c_m\log(1+n)$. $-c_m$ from data in (a) and compare with the theoretically predicted one (red curve): $-c_m=\frac{(2m-3)}{(1-m)}$. }
    \label{fig:RMT}
\end{figure}

Interestingly, numerics in figure~\ref{fig:RMT}(a) shows that the entanglement spectrum is highly none-flat. For R\'enyi entropy ($m\geq\frac{3}{2}$), $S_n^{(m)}$ decreases to minus infinity when $n$ gets large; but for $1\leq m<\frac{3}{2}$, $S^{(m)}_n$ saturate to constant negative value as $n$ gets large.

An heuristic understanding of the integral would be noticing when $n\rightarrow\infty$, the function $\left(\frac{\sin (n+1)\theta}{\sin\theta}\right)^2$ becomes a delta-function centered at $\theta=k\pi,k\in\mathbb{Z}$, i.e., we need only to consider integral inside the range $\theta \in[-\pi/(n+1),\pi/(n+1)]$:
\begin{equation}
\begin{aligned}
e^{(1-m)S_n^{(m)}}&\approx\int_{-\pi/(n+1)}^{\pi/(n+1)}d\theta \frac{2}{\pi}\sin^2\theta\left[\frac{\sin (n+1)\theta}{\sin\theta}\right]^{2m}=\int_{-\pi}^{\pi}d\varphi (n+1)^{-1}\frac{2}{\pi}\sin^2(\varphi/n)\left[\frac{\sin \varphi}{\sin\varphi/(n+1)}\right]^{2m}\\
&\approx (n+1)^{2m-3}\int_{-\pi}^{\pi}d\varphi \frac{2}{\pi}\varphi^2\left[\frac{\sin\varphi}{\varphi}\right]^{2m}\sim (n+1)^{2m-3}\\
\end{aligned}
\end{equation}
\begin{equation}
\Longrightarrow S_{n}^{(m)}\sim \frac{2m-3}{1-m}\log (n+1)+O(n^0)
\label{eq:RMT_coefficient_method1}
\end{equation}
where in the second line we consider the case when $n$ is large and approximate $\sin(\varphi/(n+1))\approx\varphi/(n+1)$. We notice that this simple calculation remarkably explains the behavior for $m>\frac{3}{2}$.

In figure~\ref{fig:RMT}(b), we compare the coefficient $c_m$ of $S_n^{(m)}\approx c_m\log n$ from data fitting and the predicted one $\frac{2m-3}{1-m}$, we see that our prediction works perfectly well for $m\gtrsim\frac{3}{2}$ and gradually fail when $m$ approaching $\frac{3}{2}$. 

\paragraph{von Neumann entropy:~$m=1,n=+\infty$~.}
To analytically capture the behavior at $m=1$, we proceed in the following. We first rewrite the von Neuman entropy in terms of KL-divergence:
\begin{equation}
S_n^{(1)}=-\int_0^{\pi}d\theta\cdot \frac{2}{\pi}\sin^2(n+1)\theta\cdot\log\left[\frac{\frac{2}{\pi}\sin^2(n+1)\theta}{\frac{2}{\pi}\sin^2\theta}\right]=-D_{KL}(P_n||P_0)
\end{equation}
where $P_n(\theta)=\frac{2}{\pi}\sin^2(n+1)\theta$ is a set of normalized probability distribution. 

The KL-divergence naturally split into two term: $S_n^{(1)}=-\tr(P_n\log P_n)+\tr(P_n\log P_0)$. The first term can be calculated exactly, which is an integral of periodic function and produces the same result for all $n$:
\begin{equation}
\label{eq:-tr(Pn log Pn)}
S_n^{(1)}\supset-\tr(P_n\log P_n)=-\int_0^{\pi}d\theta\cdot \frac{2}{\pi}\sin^2(n+1)\theta\cdot\log\left(\frac{2}{\pi}\sin^2(n+1)\theta\right)=\log(2\pi)-1,\ \forall n
\end{equation}

For the second term $-\tr(P_n\log P_0)$, in $n=+\infty$ limit, we may approximate the fast oscillating periodic function by its average value: $\frac{2}{\pi}\sin^2(n+1)\theta\approx\frac{1}{2}\frac{2}{\pi}$, and then the integral can also be calculated exactly:
\begin{equation}
S^{(1)}_{\infty}\supset\tr(P_{\infty}\log P_0)=\int_0^{\pi}d\theta\cdot \frac{2}{\pi}\cdot\frac{1}{2}\cdot\log\left(\frac{2}{\pi}\sin^2\theta\right)=-\log(2\pi)
\label{eq:tr(P_infty log P0)}
\end{equation}
Combining equation~(\ref{eq:tr(P_infty log P0)}) and~(\ref{eq:-tr(Pn log Pn)}), we have the exact result for $S^{(1)}_{\infty}$:
\begin{equation}
S^{(1)}_{\infty}=-1>-\log2
\end{equation}
It is interesting to see that in RMT limit, we can deviate from maximally mixed state at most by an amount of the same order but less than one bit.

\paragraph{von-Neumann entropy: $m=1,n<+\infty$. } It turns out the integral $\tr(P_n\log P_1)$ can be worked out exactly. We first notice the Fourier series of $\log\sin\theta$:
\begin{equation}
\log\sin\theta=-\log2-\sum_{k=1}^{+\infty}\frac{1}{k}\cos(2k\theta),\ \theta\in(0,\pi)
\end{equation}
Using this result, we have $\tr(P_n\log P_0)=-\log(2\pi)+(n+1)^{-1}$, then the von-Neumann entropy is given by:
\begin{equation}
S_n^{(1)}=-1+\frac{1}{n+1}
\end{equation}
This is consistent with the previous calculation where $S^{(1)}_{n=\infty}=-1$.
\paragraph{Set up for approximation scheme. }We see that in the above calculation at $n=\infty$, we benefit from the observation to replace $\sin^2(n+1)\theta$ by its average value $\frac{1}{2}$. In this paragraph, we want to find out the analog of this observation at large but finite $n$. The observation is that for $n\gg1$, we may approximate $P_n(\theta)$ as a simpler one $G_n(\theta)$, which we name as \textit{grating function}:
\begin{equation}
G_n(\theta)=
\begin{cases}
\frac{2}{\pi}, & \text{when }\sin^2(n+1)\theta>\frac{1}{2} \\
0, & \text{when }\sin^2(n+1)\theta<\frac{1}{2}
\end{cases}
\end{equation}
since its form looks like the transmissivity distribution of a one-dimensional optical grating with sharp edges. We denote this substitution as \textit{grating approximation}.

We immediately observe that $-\tr(P_n\log P_1)$ and $-\tr(G_n\log P_1)$ gives the same result at $n=+\infty$. Next, we are going to apply this grating approximation to estimate R\'enyi entropy at finite $n$.

\begin{figure}[t]
    \centering
    \includegraphics[width=\textwidth]{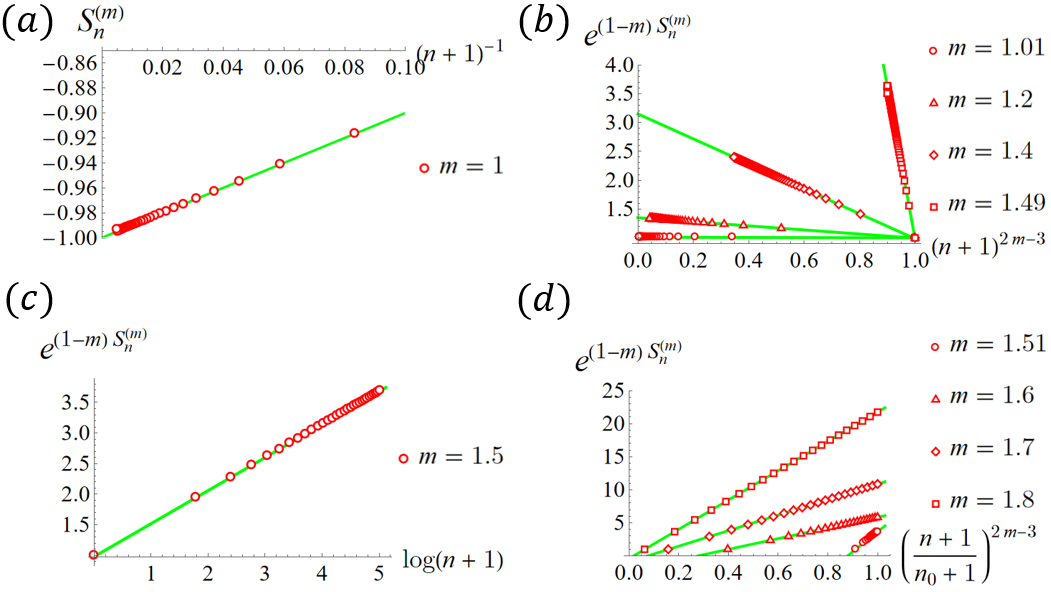}
    \caption{Linear data fitting (green curve) of R\'enyi mutual information $S_{n}^{(m)}$ in RMT limit ($q=0$). (a) $m=1$, $S_{n}^{(m)}$ is linear in $\frac{1}{n+1}$. (b) $1<m<\frac{3}{2}$, $e^{(1-m)S_{n}^{(m)}}$ is linear in $(n+1)^{2m-3}$ when $n$ gets large. (c) $m=\frac{3}{2}$, $e^{(1-m)S_{n}^{(m)}}$ is linear in $\log (n+1)$ when $n$ gets large. (b) $m>\frac{3}{2}$, $e^{(1-m)S_{n}^{(m)}}$ is linear in $(n+1)^{2m-3}$ when $n$ gets large. Here we rescaled $n$ by $n/n_{\text{max}}$ to plot four lines in one figure, where $n_{\text{max}}=100$.}
    \label{fig:linear data collapse}
\end{figure}

We notice that after replacing $P_n$ by $G_n$ in equation~(\ref{eq:1}) of R\'enyi entropy, we obtain:
\begin{equation}
S_n^{(m)}\approx\frac{1}{1-m}\log\left( \int_{0}^{\pi}d\theta\cdot G_n(\theta)\sin^{2-2m}\theta\right)
\end{equation}
where we use the fact that the value of $\frac{\pi}{2}G_n(\theta)$ is either 1 or 0, so its $2m$-th power equals itself. So, we need to develop a reasonable approximation of calculating integral like $\int_0^{\pi} d\theta\cdot G_n(\theta)f(\theta)$. We also notice that $f(\theta)$ diverges at $\theta=0,\pi$, therefore we need to treat the edge of the integral domain carefully. According to the shape of $G_n(\theta)$, the integral can be approximated as the following form when $n$ is large but finite:
\begin{equation}
\int_0^{\pi}d\theta\cdot G_n(\theta)f(\theta)\approx\frac{1}{2}\int_{\pi/4(n+1)}^{\pi-\pi/4(n+1)}d\theta\cdot\frac{2}{\pi}f(\theta)
\end{equation}
The above approximation originates from two reasons: (1) the change of integral range at the edge of domain is considered exactly (2) for the bulk of integral since $f(\theta)$ is smooth, we can well approximate the highly oscillating function (recall that we take $n$ to be large) by its average.

So the estimation of R\'enyi entropy at $n\gg1$ under grating approximation is:
\begin{equation}
S_{n}^{(m)}\approx\frac{1}{1-m}\log\left[\frac{1}{\pi}\int_{\pi/4(n+1)}^{\pi-\pi/4(n+1)}d\theta\cdot\sin^{2-2m}\theta\right]
\label{eq:grating approximation of Renyi entropy}
\end{equation}

\paragraph{R\'enyi entropy at $m>\frac{3}{2}$~.}
A first consistency check would be that: if we use the same strategy in equation~(\ref{eq:grating approximation of Renyi entropy}) at $n=\infty$, can we recover $S_{n=\infty}^{(m>1.5)}=-\infty$? The answer is indeed obviously yes for $m\geq1.5$, this is because at $\theta\rightarrow0$, the integrand $\theta^{2-2m}$ diverges.

The next step is to obtain the result when $n$ where the divergence is regulated:
\begin{equation}
\begin{aligned}
\text{Integral}&=\frac{1}{\pi}\int_{\frac{\pi}{4(n+1)}}^{\pi-\frac{\pi}{4(n+1)}}d\theta\cdot\sin^{2-2m}\theta\approx\frac{2}{\pi}\int_{\frac{\pi}{4(n+1)}}^{1}d\theta \cdot\theta^{2-2m}+\text{(regular terms)}\\
&=\frac{2}{\pi}\frac{(\pi/4(n+1))^{3-2m}}{2m-3}+\text{(regular terms)}\sim (n+1)^{2m-3}
\end{aligned}
\label{eq:RMT_coefficient_method2}
\end{equation}
where we separate the integral into its nearly divergent part and the finite part. We see that we indeed obtain and justify the same result of $S_n^{(m)}\sim \frac{2m-3}{1-m}\log (n+1)$ as in equation~(\ref{eq:RMT_coefficient_method1}). 

As a consistency check, we notice that the integral at $m=2$ can be calculated easily by noticing $\int_0^{\pi}d\theta\cdot\sin^{-2}\theta\cdot\sin^2(n\theta)=n\pi$ and rewriting $\sin^4(n\theta)=\sin^2(n\theta)-\frac{1}{4}\sin^2(2n\theta)$. The compact analytical result is given by:
\begin{equation}
S_n^{(m=2)}=-\log(n+1)
\end{equation}
which matches our prediction of scaling behaviour with respect to $n$ and even the coefficient $\frac{2m-3}{1-m}$ exactly.

\paragraph{R\'enyi entropy at $m=\frac{3}{2}$~.}
When $m=\frac{3}{2}$, we similarly have:
\begin{equation}
\begin{aligned}
\text{Integral}&=\frac{1}{\pi}\int_{\frac{\pi}{4(n+1)}}^{\pi-\frac{\pi}{4(n+1)}}d\theta\cdot\sin^{-1}\theta\approx\frac{2}{\pi}\int_{\frac{\pi}{4(n+1)}}^{1}d\theta \cdot\theta^{-1}+\text{(regular terms)}\\
&=\frac{2}{\pi}\log\left(\frac{4(n+1)}{\pi}\right)+\text{(regular terms)}\sim \log (n+1)
\end{aligned}
\end{equation}
So the R\'enyi entropy reads as: $S_{n}^{(m=1.5)}\sim-2\log\log (n+1)$, which has a slower divergence comparing to $m>\frac{3}{2}$'s case.
\paragraph{R\'enyi entropy at $1<m<1.5$~.} In this region, the integral in Equation~(\ref{eq:grating approximation of Renyi entropy}) is convergent:
\begin{equation}
\begin{aligned}
\text{Integral}&=\frac{1}{\pi}\int_{\frac{\pi}{4(n+1)}}^{\pi-\frac{\pi}{4(n+1)}}d\theta\cdot\sin^{2-2m}\theta\approx\frac{1}{\pi}\int_{0}^{\pi}d\theta\cdot\sin^{2-2m}\theta-\frac{2}{\pi}\int_{0}^{\pi/4(n+1)}d\theta \cdot\theta^{2-2m}\\
&=\frac{\Gamma(3/2-m)}{\sqrt{\pi}\Gamma(2-m)}-\frac{2}{\pi}\frac{1}{3-2m}\left(\frac{4(n+1)}{\pi}\right)^{-(3-2m)}
\end{aligned}
\end{equation}
So, the estimation for R\'enyi entropy at large $n$ would be: $S_n^{(m)}\sim -a_m+b_m (n+1)^{-(3-2m)}$ where $a_m,b_m$ is some finite positive coefficient.

In figure~\ref{fig:linear data collapse} we perform numerical simulation and careful linear data collapse to show that our prediction for the large $n$ scaling of $S_n^{(m)}$ at various ranges of $m$ is correct.

\section{SYK$_2$ limit: $q=1$}
\label{sec:q=1}
In this section, we consider a limit where $q\approx1, q^n\approx1$. In other word, if we parametrize $q=e^{-\lambda}$, we need $\lambda\rightarrow0,n\ll\lambda^{-1}$. We notice that this is not the triple scaling limit where the usual large-$p$ $\text{SYK}_p$ is recovered by requiring the scaling of $\lambda\rightarrow0, q^n\sim O(\lambda^2)$, i.e., $n= 2\lambda^{-1}\log(\lambda^{-1})+O(\lambda^{-1})$.

Instead, we observe that here we recover the limit of $\text{SYK}_2$. We first notice that in this limit,
\begin{equation}
b_n=\sqrt{\frac{1-q^n}{1-q}}=\sqrt{\frac{1-e^{-\lambda n}}{1-e^{-\lambda}}}\approx\sqrt{\frac{\lambda n}{\lambda}}=\sqrt{n}
\end{equation}
Then the symmetric transfer matrix would be $\tilde{T}=a+a^{\dagger}$ where $a^{\dagger}$ is the boson creation operator of a simple harmonic oscillator. So the moment of Hamiltonian is simplified to be:
\begin{equation}
\begin{aligned}
\mu_{2n}&=\tr H^{2n}=\langle0|(a+a^{\dagger})^{2n}|0\rangle=2^n\langle0|\hat{x}^{2n}|0\rangle\\
&=2^n\int_{-\infty}^{+\infty}dx\ x^{2n}|\psi_0(x)|^2=2^n\int_{-\infty}^{+\infty}dx\ x^{2n}\frac{1}{\sqrt{\pi}}e^{-x^2}\\
&=2^n\frac{1}{\sqrt{\pi}}\Gamma\left(n+\frac{1}{2}\right)=(2n-1)!!
\end{aligned}
\end{equation}
where we identified $\hat{x}=\frac{a+a^{\dagger}}{\sqrt{2}}$ as the position operator of the harmonic occilator and $\psi_0(x)=\langle x|0\rangle$ in the ground state wave-function in position basis. The exact form of $\mu_{2n}$ in $\text{SYK}_2$ model is calculated in appendix E of \cite{Garcia-Garcia:2018fns}, where $\mu_{2n}=(2n-1)!!+O(N^{-1})$. We see that our result matches in large $N$ limit.

Back to the theme, before calculating the $S_n^{(m)}$, we first need to prepare the form of distribution function $\Psi(\theta,q)$ and $q$-Hermite polynomial $H_n(x|q)$ in this limit.

We first consider distribution function $\Psi(\theta,q)$. Using the fact~\cite{Berkooz2018}:
\begin{equation}
\log[(e^{+2i\theta};q)_{\infty}(e^{-2i\theta};q)_{\infty}]\approx\frac{-1}{\lambda}(\text{Li}_2(e^{2i\theta})+\text{Li}_2(e^{-2i\theta}))=\frac{\pi^2}{6\lambda}-2\lambda^{-1}(\theta-\frac{\pi}{2})^2
\end{equation}
then the distribution function became a Gaussian center at $\pi/2$ with width $\varphi\equiv\theta-\frac{\pi}{2}\sim\lambda^{1/2}$:
\begin{equation}
\Psi(\theta,q)\approx\frac{C(q;q)_{\infty}e^{\frac{\pi^2}{6\lambda}}}{2\pi}\cdot e^{-2\lambda^{-1}(\theta-\frac{\pi}{2})^2}
\end{equation}
where $C$ is a constant which donot depends on $\lambda$, and is to be fixed later by normalization.

Next, we consider the approximation of $q$-Hermite polynomial. A seemingly correct reduction is by noticing that $(q;q)_k\approx \lambda^kk!$, then we have:
\begin{equation}
\begin{aligned}
H_n(x|q)&=\sum_{k=0}^{n}\frac{(q;q)_n}{(q;q)_{k}(q;q)_{n-k}}e^{i(n-2k)\theta}\approx\sum_{k=0}^{n}\frac{\lambda^nn!}{\lambda^kk!\lambda^{n-k}(n-k)!}e^{i(n-2k)\theta}\\
&=e^{in\theta}(1+e^{-i2n\theta})^n=2^n\cos^n\theta=2^nx^n
\end{aligned}
\end{equation}
Although this $2^nx^n$ is the $O(\lambda^0)$ term in the polynomial, however, this approximation is actually problematic since $x=\cos\theta=\sin\varphi\approx\varphi\sim\sqrt{\lambda}$ also has scaling dependence on $\lambda$ according the width of Gaussian distribution. For a concrete example, the $H_4(x|q)$ is given by:
\begin{equation}
\begin{aligned}
H_4(x|q)&=16x^4-4x^2(3-q-q^2-q^3)+(1-q-q^3+q^4)\\
&\approx16x^4-24x^2\lambda+3\lambda^2
\end{aligned}
\end{equation}
where in the second line we approximate the coefficient by its leading order in $\lambda$. We see that all three terms are of $O(\lambda^2)$. More generally, $H_n(x|q)$ is approximately a homogeneous function of order $O(\lambda^{n/2})$ if we consider $x\sim O(\sqrt{\lambda})$.

\begin{figure}[t]
    \centering
    \includegraphics[width=\textwidth]{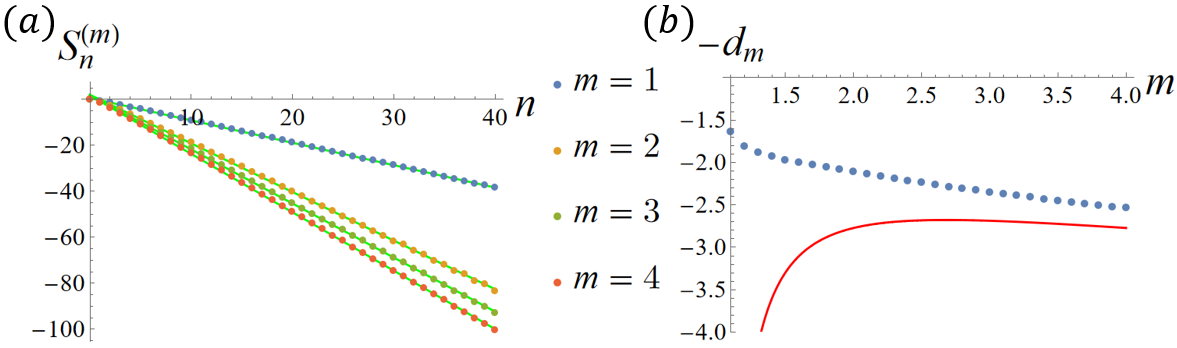}

    \caption{(a) Entropy in SYK$_2$ limit ($q=1$). The colored dots are $S_{n}^{(m)}$ from equation~(\ref{eq:2}). The solid green lines under the dots are of the ansatz form $S_{n}^{(m)}=a_m-d_m n$ and the coefficient $a_m,d_m$ are from the linear fitting of data. (b) Comparing coefficient $-d_m$ from data fitting in (a) and the theoretically estimated one: $-d_m=\frac{m\log2m}{(1-m)}$. }
    \label{fig:SYK2}
\end{figure}

To obtain the correct reduction of $q$-Hermite polynomial $H_n(x|q)$, a natural guess would be that it should be reduced to the usual Hermite polynomial $H_n(x)$, which is in wave functions of harmonic occilator since the algebraic structure already appeared in $\tilde{T}$. Actually, the answer is given by:
\begin{equation}
H_n(x|e^{-\lambda})\approx\left(\frac{\lambda}{2}\right)^{n/2}H_n\left(x/\sqrt{\frac{\lambda}{2}}\right)
\end{equation}
This can be checked directly from the recursion relation of $H_n(x|q)$ and approximate $1-q^n\approx\lambda n$, which leads to $2nH_{n-1}(y)+H_{n+1}(y)=2yH_n(y)$ and indeed confirms the recursion relation of Hermite polynomial.

From all the approximation above, we are well prepared to calculate the entanglement entropy:
\begin{equation}
\begin{aligned}
e^{(1-m)S_n^{(m)}}&\approx\frac{C(q;q)_{\infty}e^{\frac{\pi^2}{6\lambda}}}{2\pi}\int_{0}^{\pi}d\theta\ e^{-2\lambda^{-1}(\theta-\frac{\pi}{2})^2}\left[\frac{H_n^2(\cos\theta/\sqrt{\lambda/2})}{2^nn!}\right]^m\\
&\approx\frac{C(q;q)_{\infty}e^{\frac{\pi^2}{6\lambda}}}{2\pi}\sqrt{\frac{\lambda}{2}}\int_{-\infty}^{+\infty}dx\ e^{-x^2}\left[\frac{H_n^2(x)}{2^nn!}\right]^m\\
&=\frac{1}{\sqrt{\pi}}\int_{-\infty}^{+\infty}dx\ e^{-x^2}\left[\frac{H_n^2(x)}{2^nn!}\right]^m
\end{aligned}
\label{eq:2}
\end{equation}
where we used approximation~\cite{Berkooz2018,Goel2023svz}$(q;q)_{\infty}\approx\sqrt{2\pi/\lambda}\exp[-\frac{\pi^2}{6\lambda}+\frac{\lambda}{24}]$ and $C=2$ fixed by normalization of $\text{RHS}=1$ at $m=1,\forall n$. From the first line to the second, we change the dummy variable and extend the integral region to infinity.

An equivalent way of writing this would be substituting $H_n(x)$ in terms of normalized wave function $\phi_n(x)=\langle x|n\rangle$ of Harmonic oscillator:
\begin{equation}
e^{(1-m)S_n^{(m)}}=\int_{-\infty}^{+\infty}dx\  \phi_0(x)^2\left[\frac{\phi_n(x)^2}{\phi_0(x)^2}\right]^m,\ \phi_n(x)=\frac{H_n(x)e^{-x^2/2}}{\sqrt{\pi^{1/2}2^nn!}}
\end{equation}
and the normalization at $m=1$ is obvious. Similar as equation~(\ref{eq:D_KL}), the von Neumann entropy in $\text{SYK}_2$ limit also has a KL-divergence form:
\begin{equation}
S_n^{(m=1)}=-\int_{-\infty}^{+\infty}dx \phi_n(x)^2\log\left[\frac{\phi_n(x)^2}{\phi_0(x)^2}\right]=-D_{KL}(\phi_n^2||\phi_0^2)
\label{eq:D_KL harmonic}
\end{equation}
An important difference between equation~(\ref{eq:D_KL}) and equation~(\ref{eq:D_KL harmonic}) is that: for the former, $\theta$ is the indices labeling the energy eigenstates, but for the latter $n$ is the eigenstate label.

To make further analytical progress, we approximate $H_n(x)$ by its highest power $H_n(x)\approx 2^nx^n$, then the Gaussian integral can be performed and the entropy reads:
\begin{equation}
\begin{aligned}
S_n^{(m)}&\approx\frac{1}{1-m}\log\left[\frac{1}{\sqrt{\pi}}\frac{2^{mn}}{(n!)^m}\Gamma\left(mn+\frac{1}{2}\right)\right]\\
&\approx\frac{m\log2m}{1-m}n+O(\log n)
\end{aligned}
\end{equation}
where in the second line we used the Stirling formula when considering $n\gg1$.

This result hints at the linear decrease of entropy in $n$, in contrast with the $\log n$ decrease in RMT limit. As before, the coefficient in front of $n$ should not be considered to be exact since it obviously don't give a finite limit when $m\rightarrow1$. In figure~\ref{fig:SYK2}(b), we numerically confirm that the prediction of coefficient becomes better for $m$ being large. For $m$ approaching 1, the linear-in-$n$ behaviour still presence, see figure~\ref{fig:SYK2}(a).

\section{Triple scaling limit: $q\rightarrow1^{-}$}
\label{sec:triple scaling}
The usual $\text{SYK}_p$ model in large $p$ is reobtained by taking triple scaling limits:
\begin{equation}
\lambda\rightarrow0,\ n\rightarrow\infty,\ q^n=e^{-\lambda n}/\lambda^2=e^{-\tilde{\ell}}=\text{fixed}
\end{equation}
which is equivalent to say $n$ is of order $n\gtrsim 2\lambda^{-1}\log\lambda^{-1}$. Here, $\tilde{\ell}$ is the renormalized length in~\cite{Lin2022}'s bulk reconstruction in triple scaling limit. From now on we are interested in von Neumann entropy only, therefore for notational simplicity, we denote $S_n^{(m=1)}$ as $S_n$ and we will call `von Neumann entropy' simply as `entropy'.
\subsection{Intermidiate range of $q\in(0,1)$}
\label{sec:intermediate range}

\begin{figure}[t]
    \centering
    \includegraphics[width=\textwidth]{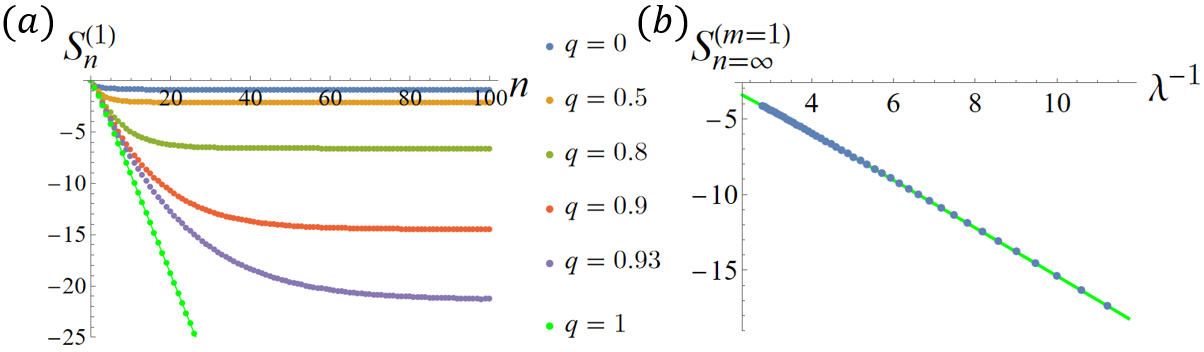}

    \caption{(a) Numerical results of von Neumann entropy $S_{n}^{(m=1)}$ at different $q$. (b) $S_{n=\infty}$ as a function of $\lambda^{-1}$(blue dots, evaluated within $q\in[0.7,0.93]$). The green line is the ansatz form $f(\lambda^{-1})=c_1\lambda^{-1}+c_2\log\lambda^{-1}+c_3$, with the coefficients from linear data fitting. We confirmed these three coefficients agree with analytical prediction in equation~(\ref{eq:Delta I_n=infty}). The maximal value of $n$ within machine precision in \texttt{Mathematica} is $n_{\text{max}}=110$ at $q=0.93$.}
    \label{fig:general q}
\end{figure}

Before diving into triple scaling limit, it is instructive to see the behavior of entanglement entropy in the intermediate range $q\in(0,1)$. Away from the two analytically controlled limits ($q=0$ or $q=1$), we only obtain some numerical results, as shown in figure~\ref{fig:general q}(a). 

We see that the $\text{SYK}_2$ limit (linear decrease with $n$) and RMT limit (remain constant) still manifest themselves in small $n$ and large $n$, respectively. Since $\text{SYK}_2$ region is only valid in $n\ll\lambda^{-1}$, we see that the initial linear decreasing region indeed expands as $q$ approaching $1$, as expected.

\subsection{Plateau value for $\lambda\rightarrow0$}
\label{sec:plateau value}
Before diving into triple scaling limit, we can first evaluate the finite saturation value of $S_n$ at $n\rightarrow \infty$ observed in figure~\ref{fig:general q}(a). This is because the probability distribution $\psi_n^2(\cos\theta|q)$ would also approach RMT value $\frac{2}{\pi}\sin^2(n+1)\theta$ even for $q\rightarrow1^{-}$, as long as $n$ is much larger than any other scales controlled by functions of $\lambda$ (see appendix \ref{appendix:edge wave function at triple scaling} for explicit justification). This point is justified in the next section when we study the low energy theory of triple scaling limit using the exact solution of Liouville quantum mechanics.

In this section, the only thing different from RMT limit is that the distribution function $P_0(\theta)=\Psi(\theta,q)$ is not $\frac{2}{\pi}\sin^2\theta$ anymore, but given by the following approximated form~\cite{Berkooz2018} in $\lambda\rightarrow0$ limit of equation~(\ref{eq:density if state}):
\begin{equation}
\Psi(\theta,q)\approx4\sqrt{\frac{2}{\pi\lambda}}e^{-2\pi^2\lambda^{-1}}e^{-2\lambda^{-1}\left(\theta-\frac{\pi}{2}\right)^2}\sin\theta\sinh\left(\frac{2\pi\theta}{\lambda}\right)\sinh\left(\frac{2\pi(\pi-\theta)}{\lambda}\right)
\end{equation}
Then, as usual, the entropy split into two terms: $S_{n=\infty}=-\tr(P_{\infty}\log P_\infty)+\tr(P_{\infty}\log P_0)$. The first term is the same as equation~(\ref{eq:-tr(Pn log Pn)}). For the second term, we need to perform the integral of $\log\Psi(\theta,q)$, where we see that different terms nicely become summations after taking logarithm, which can be integrated separately. The only non trivial integral is $\int dx \log\sinh x$, which is given in terms of polylogarithm function:
\begin{equation}
\int dx\ \log(\sinh x)=\frac{1}{2}x^2-x\log2+\frac{1}{2}\operatorname{Li}_2(e^{-2x}),\ \  \text{Li}_s(z)=\sum_{k=1}^{+\infty}\frac{z^k}{k^s}
\end{equation}
In order to calculate definite integral, one more thing needed is that $\operatorname{Li}_s(1)=\zeta(s),\ \zeta(2)=\frac{1}{6}\pi^2$.

Collecting all discussions above, we finally arrive at the explicit form of entanglement entropy at $n=\infty$:
\begin{equation}
S_{n=\infty}=-\frac{\pi^2}{6}\lambda^{-1}+\frac{1}{2}\log(\lambda^{-1})+\frac{1}{2}\log(2\pi)-1-\frac{1}{12}\lambda+\frac{\lambda}{2\pi^2}\operatorname{Li}_2(e^{-4\pi\lambda^{-1}})
\label{eq:Delta I_n=infty}
\end{equation}
where we arrange above equation in decreasing importance of $\lambda$-dependence as $\lambda^{-1}\rightarrow+\infty$.

We see that the leading order contribution is $O(\lambda^{-1})$ with a minus sign in the front. This is physical since $\lambda^{-1}=\frac{N}{2p^2}$, where $p$ is subindex of $\text{SYK}_p$ and $N$ is the number of fermions. Therefore, the leading order proportionality of $N$ means entanglement entropy is an extensive quantity that counts the degree of freedom. The first three leading order contributions ($O(\lambda^{-1}), O(\log(\lambda^{-1})),O(1)$) to plateau value is confirmed by fitting the numerical result in figure~\ref{fig:general q}(b).

We also notice that the physical entanglement entropy differs from $S_n$ by a constant shift $S_n\rightarrow S_n+S_{\text{max}}=S_n+\log\sqrt2\cdot N$. Therefore we expect that these two extensive terms should cancel each other to make the physical entanglement entropy $S_n$ be not of order $O(N)$. 

The reason why here the coefficient cannot cancel exactly is that we are not calculating a finite-$N$ theory, instead, $N$ is taken to be infinity in the first place. In a true finite $N$ calculation, one may expect that $\tr H^{2n},\ n\rightarrow\infty$ is dominated by the two eigenvectors at the edge of the spectrum (assumed to be reflection-symmetric over $H\rightarrow-H$), i.e., $\tr (H^{2n})\approx E^{2n}_{\max}|-E_{\max}\rangle\langle -E_{\max}|+E^{2n}_{\max}|E_{\max}\rangle\langle E_{\max}|$, therefore the entanglement entropy of this operator would be $\log2$, which is not extensive in $N$.

\subsection{Estimation of entropy in triple scaling limit}
\label{sec:estimation of triple scaling}

Now we are ready to calculate $S(\tl)$ at triple scaling limit. When we are seriously taking $\lambda^{-1}=0$ while keeping $\tl$ fixed, we will find that $S(\tl)$ differs from $S_{n=\infty}$ by an order one value. Therefore $S(\tl)$ itself diverges to minus infinity as $S_{n=\infty}$, so it is only meaningful to study $\Delta S(\tl)\equiv S(\tl)-S_{n=\infty}$.

In order to calculate $S(\tl)$, the qualitative feature of probabilistic distribution $P_n(\theta)=\psi^2_n(\cos\theta|q)$ is needed. The relevant low energy ($\theta\sim O(\lambda)$) behavior is worked out in appendix~\ref{appendix:edge wave function at triple scaling} using the exact solution of Liouville quantum mechanics. The transfer matrix $\tilde{T}\sim-\partial^2_{\tl}+e^{-\tl}$ describe a one-dimensional quantum mechanics of particle moving in the potential of shape $V(\tl)=e^{-\tl}$. The energy level is labeled by the asymptotic momentum $k=\theta/\lambda$, and $P_n(\theta)=|\psi_k(\tl)|^2$ is the square of eigenfunction. We denote the triple scaling limit of $P_n(\theta)$ as $P(\tl,k)$, with $k=\theta/\lambda$.

To calculate $S(\tl)$, we calculate $-\tr(P_n\log P_n)$ and $\tr(P_n\log P_0)$ respectively.

\begin{figure}[t]
    \centering
    \includegraphics[width=\textwidth]{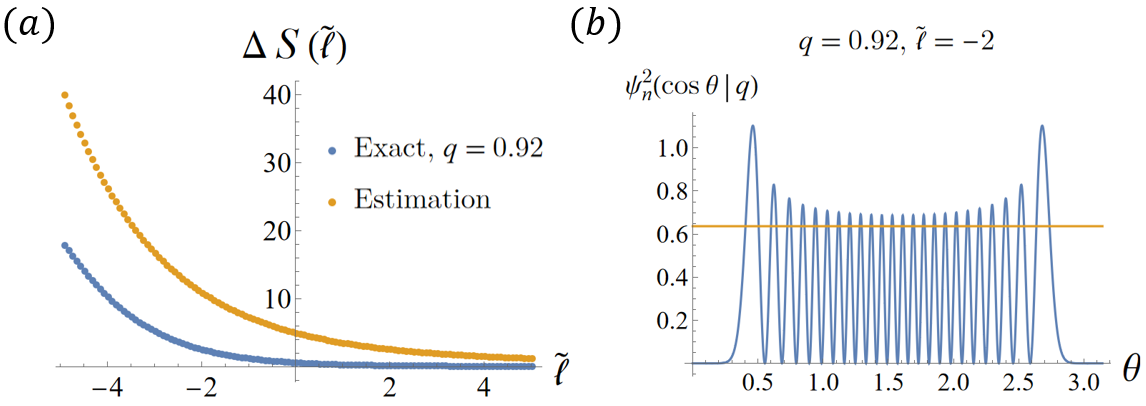}
    \caption{(a) $\Delta S(\tl)$ as a function of $\tl$, comparing the numerical result at finite $\lambda$ with estimation $\Delta S(\tl)\sim\pi k_0(\tl)$. (b) A typical shape of $\psi^2_n(\cos\theta|q)$. The orange horizon line is $\frac{2}{\pi}$.}
    \label{fig:triple scaling}
\end{figure}

First, we show that $-\tr(P_n\log P_n)$ term is the same as $n=\infty$ result, which is given by equation~(\ref{eq:-tr(Pn log Pn)}). This is because in triple scaling limit, $P_n(\theta)\approx\frac{2}{\pi}\sin^2(n\theta)$ for $\theta\sim O(1)$ (see appendix~\ref{appendix:edge wave function at triple scaling} and figure~\ref{fig:triple scaling}(b)), they only differ at the edge of integral domain $\theta\in(0,\pi)$, where $\theta\sim O(\lambda)$ or $(\pi-\theta)\sim O(\lambda)$. Since $P_n(\theta)$ is everywhere bounded when $\lambda\rightarrow0$, we conclude that $-\tr(P_n\log P_n)-(-\tr(P_{\infty}\log P_{\infty}))\sim O(\lambda)$, which vanishes in strict triple scaling limit.

So, the contribution of entropy difference solely comes from $\tr((P_n-P_\infty)\log P_0)$ terms:
\begin{equation}
\begin{aligned}
\Delta S(\tl)&=\lim\limits_{\lambda\rightarrow0}2\int_0^{\frac{\pi}{2}}d\theta\left(P_n(\theta)-\frac{1}{\pi}\right)\log\Psi(\theta,q)\\
&=2\int_0^{+\infty}dk\left(P(\tl,k)-\frac{1}{\pi}\right)\cdot\left[\lim\limits_{\lambda\rightarrow0}\left(\lambda\log\Psi(k\lambda,e^{-\lambda})\right)\right]
\end{aligned}
\end{equation}
Here, the factor $2$ in front of the integral is because of the inversion symmetry of integrand over $\theta\rightarrow\pi-\theta$. To perform the actual calculation, we first need the distribution function $\Psi(\theta,q)$ in low energy limit ($\theta\sim O(\lambda)$)~\cite{Berkooz2018}:
\begin{equation}
\Psi(\theta,q)=2\sqrt{\frac{2}{\pi\lambda}}e^{-\frac{1}{2}\pi^2\lambda^{-1}}e^{-2\lambda^{-1}\theta^2}\sin\theta\sinh\left(\frac{2\pi\theta}{\lambda}\right)
\end{equation}
In terms of variable $k$ ($k=\frac{\theta}{\lambda}\sim O(1)$), we have:
\begin{equation}
\log\Psi(k\lambda,e^{-\lambda})=\log\left(2\sqrt{\frac{2}{\pi}}\right)-\frac{1}{2}\log\lambda^{-1}-\frac{1}{2}\pi^2\lambda^{-1}-2\lambda k^2+\log k+\log\sinh(2\pi k)
\end{equation}
Therefore we observe that:
\begin{equation}
\lim\limits_{\lambda\rightarrow0}\lambda\log\Psi(k\lambda,e^{-\lambda})=-\frac{1}{2}\pi^2
\end{equation}
So the final result of entropy in triple scaling limit is given by:
\begin{equation}
\Delta S(\tl)=-\pi^2\int_{0}^{+\infty}dk\left(P(\tl,k)-\frac{1}{\pi}\right)
\label{eq:entropy at triple scaling}
\end{equation}
where $P(\tl,k)$ is given by Bessel function of first kind $K_{\nu}(z)$ (see appendix~\ref{appendix:edge wave function at triple scaling} for derivation):
\begin{equation}
P(\tl,k)=\frac{2/\pi}{|\Gamma(2ik)|^2}K_{2ik}^2(2e^{-\tl/2})
\end{equation}

The behavior of $P(\tl,k)$ is easy to understand from a semi-classical point of view. For a fixed $\tl$, there exists a typical energy scale, the `penetration energy' $k_0(\tl)$, operationally defined as the energy $k_0$ at which the probability function $P(\tl,k)$ arrive at its first maximal when increasing $k$ from zero. Physically, having WKB approximation in mind, when $k\lesssim k_0(\tl)$, $P(\tl,k)$ is almost zero in the classical forbidden region; when $k\gtrsim k_0(\tl)$, the potential energy $V(\tl)$ is much smaller than the total energy $k^2$, therefore the particle is almost free and its wavefunction is almost a standing wave: $P(\tl,k)\approx \frac{2}{\pi}\sin^2(k\tl)$. This is the reason why the integral~(\ref{eq:entropy at triple scaling}) converges at $k\rightarrow+\infty$. A justification of this argument by WKB approximation is compared in appendix~\ref{appendix:edge wave function at triple scaling}.

According to the analysis of behavior of $P(\tl,k)$, we may perform the following grating approximation by dividing the integral domain into two regions $(0,k_0(\tl))$ and $(k_0(\tl),+\infty)$, and approximate $P(\tl,k)\approx0$ in the first region and $P(\tl,k)\approx\frac{2}{\pi}\sin^2(k\tl)$ in the second. Therefore the integral in the second region vanishes, leaving a positive contribution from the first: region:
\begin{equation}
\Delta S(\tl)\approx-\pi^2\int_0^{k_0(\tl)}dk\left(0-\frac{1}{\pi}\right)=\pi k_0(\tl)
\end{equation}
In appendix~\ref{appendix:edge wave function at triple scaling} we show the behaviour of $k_0(\tl)$: (1), $k_0(\tl)\approx e^{-\tl/2}$ when $\tl\lesssim0$. This is expected from WKB approximation where the penetration energy is determined classically ($k_0^2-V(\tl)=0$, where the classical kinetic energy is zero). (2), $k_0(\tl)\approx \frac{\pi}{2}\tl^{-1}$ when $\tl\gtrsim0$. This is anticipated from the large-$\tl$ behavior of RMT result. In figure~\ref{fig:triple scaling}(a) we compare our approximation with the numerical result (we can't do numerical integral at exact triple scaling limit with $\lambda=0$, due to fast oscillation of integrand. We calculate at reasonably small $\lambda\sim0.1$). We see that this crude approximation captures the qualitative feature of $\Delta S(\tl)$, while quantitatively there are still discrepancies. Such discrepancy may come from the finite-$\lambda$-effect of numerical simulation or the non-negligible contribution from the intermediate region ($k$ is in the vicinity of $k_0(\tl)$) which glues the classically forbidden region and free particle region, where the wave function has a high peak (see figure~\ref{fig:triple scaling}(b) and figure~\ref{fig:penetrating energy}(a)).

\subsection{Matching semi-classical calculation in JT gravity}
\label{sec:matching JT}
In this section, we show that the entropy $\Delta S(\tl)$ in semiclassical region $\tl\lesssim0$ where WKB approximation of wave function works, the result $\Delta S(\tl)\sim k_0(\tl)\approx  e^{-\tl/2}$ matches the semi-classical calculation in JT gravity, where the entropy of geodesic wormhole is given by the classical solution of dilaton value at the center of the geodesic.

The JT gravity is a dilaton-gravity model on a (1+1) dimensional asymptotic AdS$_2$ manifold $\mathcal{M}$ with following Lorenzian action:
\begin{equation}
S_{\text{JT}}=\Phi_0\chi_{\text{Euler}}+\int_{\mathcal{M}}\sqrt{-g}\Phi(R+2)+2\int_{\partial\mathcal{M}}\sqrt{-\gamma}\Phi(K-1)
\end{equation}
The first term $\chi_{\text{Euler}}$ is the Eintein-Hilbert action with proper Gibbons-Hawking-York boundary term which together composed of Euler character. This purely topological, though do not contribute to dynamics, has a significant role in the topological expansion of JT gravity as a matrix integral~\cite{Saad:2019lba}. The second term is the bulk dilaton action with linear dilaton potential. Integrating over $\Phi$, we set Ricci scalar $R=-2$, resulting in a rigid bulk AdS$_2$ spacetime. Variation over bulk metric we can get the on-shell equation-of-motion of dilaton field. The non-trivial dynamics happen at the left and right timelike boundary, where the holographic boundary observer lives.

\begin{figure}[t]
    \centering
    \includegraphics[width=0.25\textwidth]{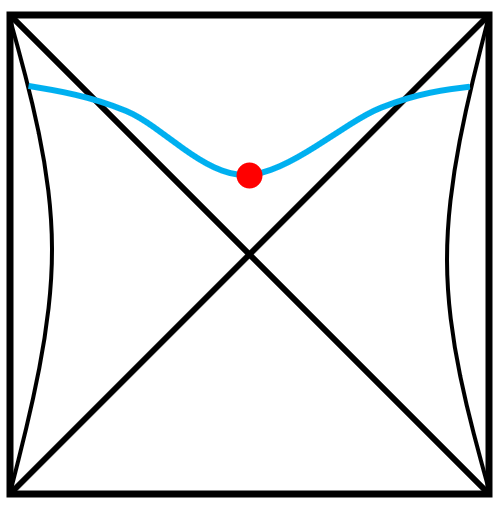}
    \caption{Geodesic wormhole (blue curve) in AdS$_2$ connecting left and right asymptotic boundaries. The red dot emphasizes that the entropy of this wormhole state is given by the value of dilaton at the center.}
    \label{fig:enter-label}
\end{figure}

In this section, we will first work out the classical solution of metric and dilaton field, then calculate the geodesic length connecting the left and right asymptotic boundary. According to the RT formula, the entropy is proportional to the extremal value of area (codimension 2 sub-manifold) that is homologous to a boundary region. In dilaton gravity in 1+1 dimension, the value of area is replaced by the dilaton value. (This can be interpreted from the fact that JT gravity originates from the $s$-wave reduction of higher dimensional near extremal black hole~\cite{Mertens2022}, where the dilaton is the fluctuation of higher dimensional area.) Therefore, due to the reflection symmetry between left/right, the extremal dilaton profile lies at the center of the geodesic. See figure~\ref{fig:enter-label} for illustration.

\paragraph{Classical solution of metric.} AdS$_2$ can be embedded in a 3-dimensional Minkovski space with signature $(-,-,+)$, with metric:
\begin{equation}
ds^2=-dT_1^2-dT_2^2+dX^2
\end{equation}
AdS$_2$ is the universal cover of the induced metric on the hyperbola:
\begin{equation}
    T_1^2+T_2^2-X^2=1
\end{equation}
We will be interest in two kinds of parametrization of hyperbola: (1) Rindler coordinate, which is useful to place the asymptotic boundary and match the boundary clock for holographic observer; (2) global coordinate, which is useful to calculate the the geodesic length and read out the value of dilaton at the center of geodesic, whose location extends into the region which Rindler coordinates cannot cover.

The two coordinate systems are given by:
\begin{equation}
\text{global: }\left\{
\begin{aligned}
&T_1=\sqrt{1+x^2}\cos t_g\\
&T_2=\sqrt{1+x^2}\sin t_g\\
&X=x\\
\end{aligned}
\right.
,\ \ \text{Rindler: }
\left\{
\begin{aligned}
&T_1=r/r_h\\
&T_2=\sqrt{(r/r_h)^2-1}\sinh(r_ht)\\
&X=\sqrt{(r/r_h)^2-1}\cosh(r_ht)\\
\end{aligned}
\right.
\end{equation}
with induced metric:
\begin{equation}
ds^2=-(1+x^2)dt_g^2+\frac{dx^2}{1+x^2},\ \ ds^2=-(r^2-r_h^2)dt^2+\frac{dr^2}{r^2-r_h^2}
\end{equation}
The subscript `$g$' is for `global' and `$h$' for `horizon' ($r=r_h$ is Rindler horizon). The domain of Rindler coordinate is $r>r_h, t\in(-\infty,+\infty)$, which covers the right Rindler patch of figure~\ref{fig:enter-label}. The corresponding domain of global coordinate is $x\in(-\infty,+\infty), t_g\in[-\pi/2,\pi/2]$.

\paragraph{Boundary and boundary conditions. }The right asymptotic boundary is placed at $r=r_b$, with $r_b\gg r_h$ (In the end of calculation, we are taking $r_b\rightarrow+\infty$). In global coordinate, boundary place $r=r_b$ is translated into $r_b/r_h=\sqrt{1+x_b^2}\cos t_g$The boundary metric is given by $ds^2=-r_b^2dt^2$. So, we see that by dropping the constant Weyl factor, $t$ becomes the proper time for the holographic observer. The boundary condition of dilaton is that it remains constant at the boundary: $\Phi|_{\partial\mathcal{M}}=\phi_br_b$.
\paragraph{Classical solution of dilaton field.}
The general solution for dilaton field is $\Phi=AT_1+BT_2+CX$, where the SO(2,1) covariance is manifest. In our case, the physical solution is by choosing $B=C=0$~\cite{Harlow:2018tqv}, in order to match the boundary condition. So, we have dilaton profile in both coordinates:
\begin{equation}
\Phi=\Phi_h\sqrt{1+x^2}\cos t_g=\Phi_h(r/r_h)
\end{equation}
where $\Phi_h>0$ is the dilaton value on the horizon. Matching the boundary condition, we have $\phi_b=\Phi_h(r_b/r_h)$.
\paragraph{Geodesic length.}In pure JT gravity without matter, the classical ADM energy on the left/right boundary is the same~\cite{Harlow:2018tqv}: $E_L=E_R=\Phi^2_h/\phi_b$. Therefore upon covariant quantization~\cite{Harlow:2018tqv}, $H_L-H_R=0$ should be considered as a gauge constraint, namely the physical states satisfy $(H_L-H_R)|\psi\rangle=0$. Therefore the evolution of boost time should be identified as the same states: $e^{i(H_L-H_R)t}|\psi\rangle=|\psi\rangle$, indicates that the we can set $t_L=t_R\equiv t$ as a gauge fixing condition. Therefore it is sufficient to consider the geodesic with the same boundary time at left/right end point.

Now, we are ready to calculate geodesic length~\cite{Harlow:2018tqv}. In global coordinate with time translation symmetry, the $t_g=\text{Const}$ line is obviously a spacelike geodesic. The bare length is given by:
\begin{equation}
\ell_{\text{bare}}=\int_{-x_b}^{x_b}dx\cdot\frac{1}{\sqrt{1+x^2}}=2\log\left(x_b+\sqrt{1+x_b^2}\right)\approx2\log(2x_b)
\end{equation}
where we notice that $x_b\gg1$. Since $x_b$ is implicitly dependent on global time $t_g$ via $r_b/r_h=\sqrt{1+x^2_b}\cos t_g$, the last thing to do is to relate the boundary to global time: via $T_2/T_1=\frac{\sqrt{(r_b/r_h)^2-1}}{(r_b/r_h)}\sinh(r_ht)=\tan t_g$. Since $r_b/r_h\gg1$, we have $\cos t_g\approx(\cosh r_ht)^{-1}$. Therefore we obtain the bare length as a function of boundary time:
\begin{equation}
\ell_{\text{bare}}(t)=2\log\left(\frac{\cosh r_ht}{\Phi_h}\right)+2\log\left(2\phi_br_b\right)
\end{equation}
We define the renormalized length $\tl$ by dropping the universal divergent constant $2\log(2\Phi|_{\partial\mathcal{M}})=2\log(2\phi_br_b)$, which is finite in $r_b\rightarrow+\infty$ limit:
\begin{equation}
\tl(t)=2\log\left(\frac{\cosh r_ht}{\Phi_h}\right)
\end{equation}
\paragraph{Matching entropy with boundary calculation.}
The center of geodesic is located at $x=0$, so the dilaton value at the center is given by: $\Phi_{\text{center}}=\Phi_h\cos t_g=\Phi_h(\cosh r_ht)^{-1}$. Therefore we have:
\begin{equation}
\tl=2\log(\Phi_{\text{center}}^{-1})\longrightarrow S(\tl)\propto \Phi_{\text{center}}=e^{-\tl/2}
\end{equation}
We observe that the scaling behavior matches the boundary calculation of $\Delta S(\tl)$ in semi-classical region.

\section{Conclusions}
\label{label:conclusion}

In this work, we study the entanglement entropy 
and its R\'enyi generalization of fixed-length states in 0-particle Hilbert space of DSSYK model. We show that the entanglement entropy is negative, which is a manifestation of type II$_1$ von Neumann algebra. In RMT limit ($q=0$), we find that $S_n=-1+\frac{1}{n+1}$, meaning that the long wormhole cannot deviate from maximally mixed state even by one bit. We also observe that for different $m$, the scaling behavior of $S_n^{(m)}$ at large $n$ is qualitatively different. This provides an example where R\'enyi entropy is not a good correlation measure when entanglement spectrum is highly non-flat. In SYK$_2$ limit, we find that all $S_n^{(m)}$ linearly decrease with $n$. Then we dive into triple scaling limit and estimate the entanglement entropy using low energy effective theory described by Liouville quantum mechanics. In semi-classical regime, we match the estimated entropy from faithful boundary calculation to the bulk calculation in JT gravity, where the entropy is given by the on-shell value of dilaton field at the center of the geodesic, as predicted by RT formula.

An interesting future direction would be how to better interpret this boundary calculation in terms of the bulk gravity picture. For example, we show that the dilaton emerges on-shell. It would be interesting if we could see dilaton emerges from off-shell boundary calculation.

\appendix
\section{Low energy wave function at triple scaling limit}
\label{appendix:edge wave function at triple scaling}
\subsection{Liouville quantum mechanics}
In this section, we calculate the wave function $\psi_n(\cos\theta|q)$ at triple scaling limit through Liouville quantum mechanics.

From the recursion relation of the symmetrized transfer matrix $\tilde{T}$ in equation~(\ref{eq:tranfer matrix}):
\begin{equation}
(\tilde{T}\psi)_n=\sqrt{\frac{1-q^n}{1-q}}\psi_{n-1}+\sqrt{\frac{1-q^{n+1}}{1-q}}\psi_{n+1}
\end{equation}
we can search for a continuous version to represent operator $\tilde{T}$. In triple scaling limit where $q\rightarrow1^{-},q^{n}\sim O(\lambda^2)$, we approximate $\sqrt{1-q}\approx\sqrt{\lambda},\sqrt{1-q^{n+1}}\approx\sqrt{1-q^n}\approx1-\frac{1}{2}q^{n}$ and $\psi_{n+1}+\psi_{n-1}-2\psi_{n}\approx\partial_n^2\psi$.
The wormhole length is defined by $q^{-n}=e^{-\lambda\ell}$ in the main text, so the continuous version of the operator $\tilde{T}$ is given by:
\begin{equation}
\sqrt{\lambda}\tilde{T}=\lambda^2\partial_{\ell}^2-e^{-\ell}+2
\end{equation}

The triple scaling limit is obtained by absorbing the `effective $\hbar^2$', which is $\lambda^2$ in front of kinetic energy, into the redefinition of wormhole length, i.e., the \textit{renormalized} wormhole length $\tl$:
\begin{equation}
\tl\equiv \ell+2\log{\lambda}
\end{equation}
Since the spectrum of $\tilde{T}$ has reflection symmetry over zero, we define the Hamiltonian to be ($-\tilde{T}$) to make the sign of kinetic term to be minus:
\begin{equation}
-\tilde{T}=E_0+\lambda^{3/2}\left(-\partial_{\tl}^2+e^{-\tl}\right),\ E_0=\frac{-2}{\sqrt{\lambda}}
\end{equation}

We will see that this Liouville operator $\mathcal{L}\equiv\left(-\partial_{\tl}^2+e^{-\tl}\right)$ describes the low energy wave function and spectrum. A quick consistency check is by the observation that $\mathcal{L}$ has a non-negative spectrum since it describes the one-dimensional particle moving in a positive potential $V(\tl)=e^{-\tl}$, whose eigenstates are scattering states which are asymptotically free particle at $\tl\gg1, V(\tl)\rightarrow0^{+}$. Therefore the groundstate of $-\tilde{T}$ is given by $\mathcal{L}=0$, with $-\tilde{T}=E_0$. This matched the exact spectrum $-\tilde{T}=\frac{-2\cos\theta}{\sqrt{\lambda}}$ at $\theta=0$.

The eigenstates of Liouville operator $\mathcal{L}$ can be solved exactly:
\begin{equation}
\left(-\partial_{\tl}^2+e^{-\tl}\right)\psi_{k}(\tl)=k^2\psi_k(\tl), \ \psi_{k}(\tl)=\frac{2/\sqrt{2\pi}}{\Gamma(2ik)}K_{2ik}(2e^{-\tl/2})
\end{equation}
where $K_{\nu}(z)$ is the modified Bessel function of the second kind, explicitly \texttt{BesselK[$\nu$,z]} in \texttt{Mathematica}. The choice of normalization factor would be self-obvious later.

By matching the energy spectrum $-E$ of $-\tilde{T}$, we can relate parameter $k$, which is the momentum of the one-dimensional particle, to the original spectrum parameter $\theta$:
\begin{equation}
-E(\theta)=\frac{-2\cos\theta}{\sqrt{1-q}}\approx\frac{-2(1-\theta^2/2)}{\sqrt{\lambda}}=E_0+\lambda^{-1/2}\theta^2=E_0+\lambda^{3/2}k^2
\end{equation}
\begin{equation}
\Longrightarrow k=\frac{\theta}{\lambda}
\end{equation}    
From above we see that the triple scaling limit describes the low energy region where $\theta\sim O(\lambda)\ll1$ since we assume $\tl,k\sim O(1)$.

\begin{figure}[t]
    \centering
    \includegraphics[width=\textwidth]{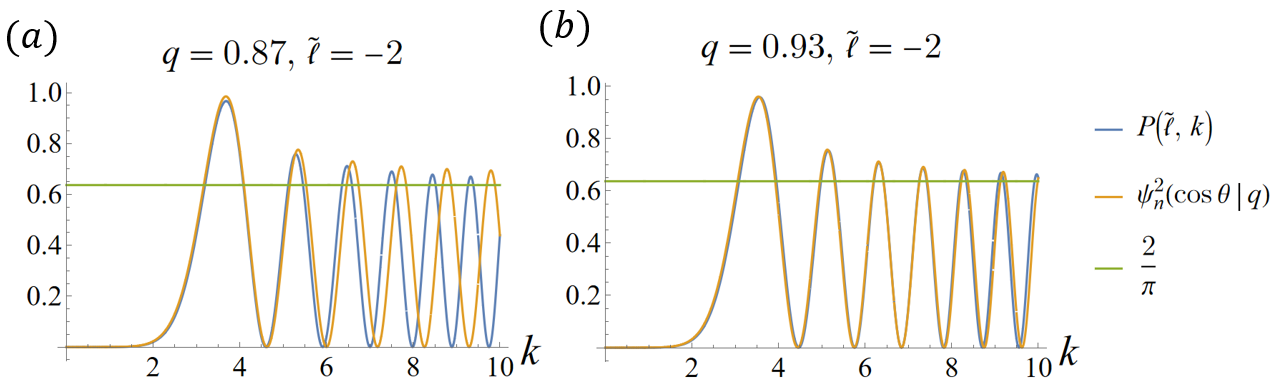}
    \caption{Compare probability distribution from low energy Liouville quantum mechanics $P(\tl,k)$ and exact result $\psi^2_n(\cos\theta|q)$ with $k=\frac{\theta}{\lambda}$. We see that from (a) to (b) as $\lambda^{-1}$ increase and $\tl$ kept fixed, low energy effective theory gradually matches .}
    \label{fig:compare Liouville and Q-pochhammer}
\end{figure}

The distribution function $P(\tl,k)=|\psi_{k}(\tl)|^2$ given by:
\begin{equation}
P(\tl,k)=\frac{2/\pi}{|\Gamma(2ik)|^2}K^2_{2ik}(2e^{-\tl/2})
\end{equation}
would then  exactly matches the exact result $\psi_{n}^2(\cos\theta|q)$ in triple scaling limit, which we show numerically in figure~\ref{fig:compare Liouville and Q-pochhammer}.

\subsection{Approaching RMT limit}
We first analyze how the distribution function $P(\tl,k)$ smoothly approaches the RMT result $\frac{2}{\pi}\sin^2n\theta$ when we further consider $\tl\gg1$. Then the argument of Bessel function is $2e^{-\tl/2}\ll1$. Using the expansion of $K_{\nu}(z)$ at small $z$:
\begin{equation}
\begin{aligned}
K_{\nu}(z)&=\frac{1}{2}\Gamma(\nu)\left(\frac{z}{2}\right)^{-\nu}\sum_{k=0}^{+\infty}\frac{(z/2)^{2k}}{(1-\nu)_{k}k!}+\frac{1}{2}\Gamma(-\nu)\left(\frac{z}{2}\right)^{\nu}\sum_{k=0}^{+\infty}\frac{(z/2)^{2k}}{(1+\nu)_{k}k!}\\
&\approx\frac{1}{2}\Gamma(\nu)\left(\frac{z}{2}\right)^{-\nu}+\frac{1}{2}\Gamma(-\nu)\left(\frac{z}{2}\right)^{\nu},\ |z|\ll1
\end{aligned}
\end{equation}
We notice that in our case $\nu=2ik$ is purely imaginary, and since $\Gamma(z^*)=(\Gamma(z))^*$, meaning that $K_{\nu}(z)\in\mathbb{R}$ for $z\in\mathbb{R}_{+}$. In this way, the distribution function is given by:
\begin{equation}
P_{\text{RMT}}(\tl,k)=\frac{2}{\pi}\cos^2\left[k\tl+\operatorname{Im}\log\Gamma(2ik)\right]
\end{equation}
For sufficiently large $\tl$ ($n$ is larger than any other scales controlled by functions of $\lambda$) and non-zero $k$, the phase is dominated by the first term with fast occilation $k\tl=\theta(n-2\lambda^{-1}\log\lambda^{-1})\approx \theta n$, which means $P(\tl,k)\approx \frac{2}{\pi}\cos^2n\theta$. 

To compare with RMT result $\frac{2}{\pi}\sin^2n\theta$, the extra phase $\frac{\pi}{2}$ shows up when we consider the correction from $\operatorname{Im}\log\Gamma(2ik)$, which is important when $k$ approaches zero. This is done by utilizing the expansion of $\log\Gamma(z)$ at small $z$:
\begin{equation}
\log\Gamma(z)=\log\Gamma(1+z)-\log z=-\log z-\gamma z+\sum_{k=2}^{+\infty}\frac{\zeta(k)}{k}(-z)^k,\ |z|<1
\end{equation}
where $\gamma\approx0.57$ is the Euler constant. Then the distribution function is expanded in small $k$:
\begin{equation}
\begin{aligned}
P_{\text{RMT}}(\tl,k)&=\frac{2}{\pi}\cos^2\left[k\tl-\frac{1}{2}\pi-2\gamma k+\sum_{p=1}^{+\infty}\frac{\zeta(2p+1)}{2p+1}(-1)^{p+1}(2k)^{2p+1}\right],\ 0<k<1/2\\
&=\frac{2}{\pi}\cos^2\left[k\tl-\frac{1}{2}\pi-2\gamma k+O(k^3)\right]=\frac{2}{\pi}\sin^2\left[k\tl-2\gamma k+O(k^3)\right]\\
&=\frac{2}{\pi}\sin^2\left[\theta(n-2\lambda^{-1}\log{\lambda^{-1}}-2\gamma\lambda^{-1})+O\left(\left(\theta/\lambda\right)^3\right)\right]
\end{aligned}
\label{eq:B.11}
\end{equation}
which explicitly matches the RMT result and provides the new information as correction.

For later convenience, it is instructive to study the low energy behavior of $P(\tl,k)$ for a fixed $\tl$. One characteristic is the \textit{penetrating energy} $k_0(\tl)$, operationally defined as the energy $k_0$ at which the probability function $P(\tl,k)$ arrive at its first maximal when increasing $k$ from $0$. Interpretation of penetrating energy $k_0(\tl)$ is obvious in view of WKB-approximation, where, semi-classically, the particle cannot go deeper into the region $\tl'<\tl$ if its energy is smaller than $k_0(\tl)$. In the view of WKB-approximation, the wave function starts to oscillate only when $k>k_0$, and is approximately zero when $k<k_0$.

In $\tl\gg1$ limit, the penetrating energy $k_0(\tl)$ is determined when the phase inside $\cos^2[...]$ increases (from $-\pi/2$) to $0$, i.e.:
\begin{equation}
k_0(\tl)\tl+\operatorname{Im}\log{\Gamma(2ik_0(\tl))}=0
\end{equation}
which is then solved perturbatively using expansion in the first line of Equation~(\ref{eq:B.11}) for $\tl\gg1$:
\begin{equation}
k_0(\tl)\approx\frac{\pi}{2}\frac{1}{\tl-2\gamma}-\frac{\pi^3\zeta(3)}{3}\frac{1}{(\tl-2\gamma)^4}+\frac{\pi^5\zeta(5)}{5}\frac{1}{(\tl-2\gamma)^6}+O\left((\tl-2\gamma)^{-7}\right)
\end{equation}

\begin{figure}[t]
    \centering
    \includegraphics[width=\textwidth]{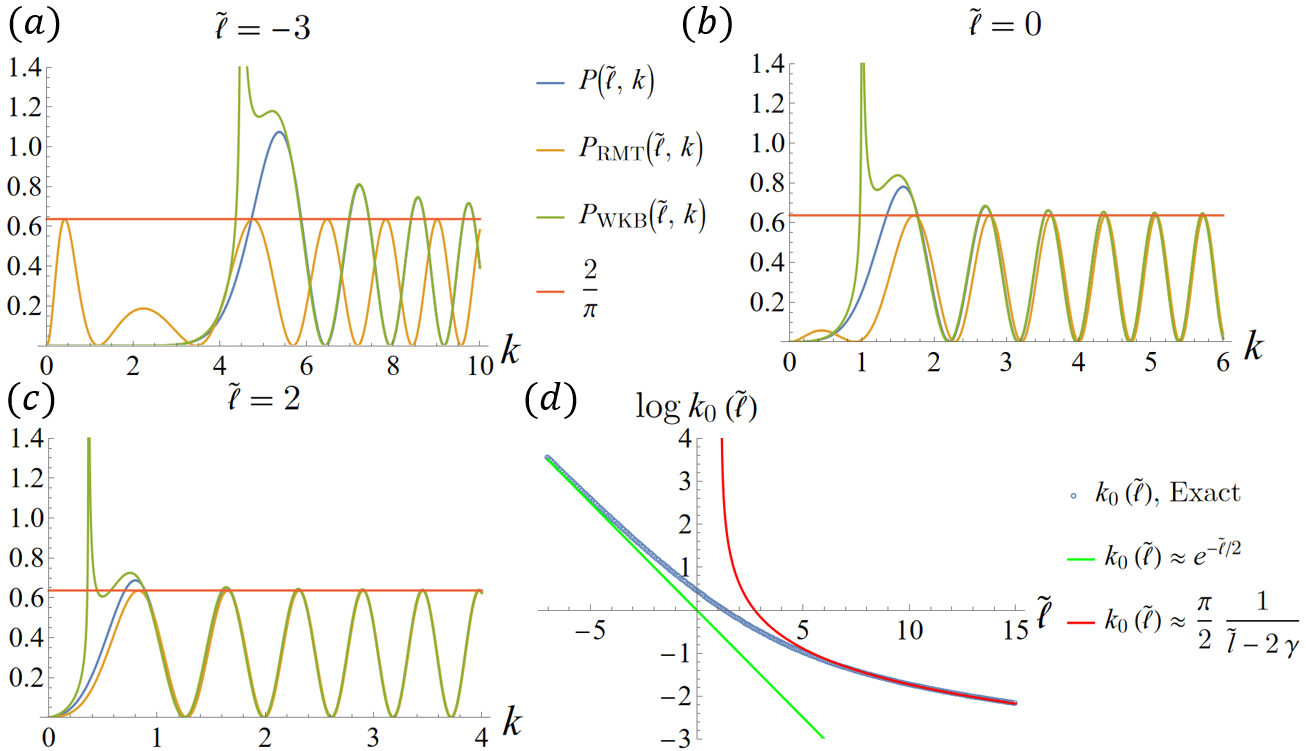}
    \caption{(a), (b), (c) Compare distribution function $P(\tl,k)$ with its approximation $P_{\text{RMT}}(\tl,k),P_{\text{WKB}}(\tl,k)$ as a function of $k$ at different $\tl$. (d) Compare the exact result of penetrating energy $k_0(\tl)$ with its approximations in different regions.}
    \label{fig:penetrating energy}
\end{figure}
\subsection{WKB limit}

Another instructive limit is $\tl<0,|\tl|\gg1$, or simply denoted as $\tl\ll-1$. More precisely, we consider the original length $\ell$ to be order one. The Liouville operator in this limit is given by $\lambda^2\mathcal{L}=-\lambda^2\partial_{\ell}+e^{-\ell}$. The effective-$\hbar$ is recovered in front of kinetic energy and a systematic semiclassical expansion can be performed, which is known as WKB approximation.

In WKB approximation, the penetrating energy $k_0(\tl)$ is obtained at vanishing kinetic energy, where the semi-classical particle is static instantaneously. Therefore acquires a high probabilistic density there. In this way, we have:
\begin{equation}
k_0(\tl)\approx e^{-\tl/2},\ \text{when}\ \tl\ll-1
\end{equation}
The qualitatively different behaviour of $k_0(\tl)$ at $\tl\ll-1$ and $\tl\gg1$ is verified numerically in figure~\ref{fig:penetrating energy}(d).

Within the scheme of WKB approximation, the probability distribution function (square of wave function) is calculated as a standard exercise of undergraduate quantum mechanics course:
\begin{equation}
P_{\text{WKB}}(\tl,k)=
\begin{cases}
\sqrt{\frac{k^2}{k^2-e^{-\tl}}}\frac{2}{\pi}\sin^2\left[\int_{\tl_0}^{\tl}d\tl'\sqrt{k^2-e^{-\tl'}}+\frac{\pi}{4}\right], & \tl>\tl_0=-2\log k\\
\sqrt{\frac{k^2}{e^{-\tl}-k^2}}\frac{1}{2\pi}\exp\left[-2\int_{\tl}^{\tl_0}d\tl'\sqrt{e^{-\tl'}-k^2}\right], & \tl<\tl_0=-2\log k
\end{cases}
\end{equation}

In figure~\ref{fig:penetrating energy}(a),(b),(c), we compare $P_{\text{WKB}}(\tl,k)$ with exact wave function $P(\tl,k)$ as a function of $k$ in different range of $\tl$. We see that WKB-approximation works well when $k$ is sway from $k_0(\tl)$.

Since we are not aware of how to do the asymptotic expansion of $K_{\nu}(z)$ in large-$\nu$ limit directly, the WKB approximation provides valuable insight. In large-$k$ limit for any $\tl$, we see that the $P(\tl,k)\approx P_{\text{WKB}}(\tl,k)\approx\frac{2}{\pi}\sin^{2}(k(\tl-\tl_0))$, i.e., the RMT result emerge for arbitrary $\tl$ as long as $k$ is large. This is reasonable as expected, and serves as a starting point of the estimation procedure in triple scaling limit applied in the main text.

\acknowledgments
I would thank Yiming Chen, Henry Lin, Xiao-Liang Qi, Xiaoyi Shi, Herman Verlinde, Jinzhao Wang, and Jiuci Xu for their helpful discussions. I would like to thank Professor Xiao-Liang Qi for encouraging me to publish these results. Part of this work was finished when I was studying at Institute for Advanced Study, Tsinghua University. I would like to thank Professor Hui Zhai for his valuable instruction and support.

\bibliographystyle{JHEP}
\bibliography{biblio}

\providecommand{\href}[2]{#2}\begingroup\raggedright\begin{thebibliography}{10}

\bibitem{Witten:2018zxz}
E.~Witten, \emph{{APS Medal for Exceptional Achievement in Research: Invited
  article on entanglement properties of quantum field theory}},
  \href{https://doi.org/10.1103/RevModPhys.90.045003}{\emph{Rev. Mod. Phys.}
  {\bfseries 90} (2018) 045003}
  [\href{https://arxiv.org/abs/1803.04993}{{\ttfamily 1803.04993}}].

\bibitem{Witten:2021jzq}
E.~Witten, \emph{{Why Does Quantum Field Theory In Curved Spacetime Make Sense?
  And What Happens To The Algebra of Observables In The Thermodynamic Limit?}},
   \href{https://arxiv.org/abs/2112.11614}{{\ttfamily 2112.11614}}.

\bibitem{Witten:2021unn}
E.~Witten, \emph{{Gravity and the crossed product}},
  \href{https://doi.org/10.1007/JHEP10(2022)008}{\emph{JHEP} {\bfseries 10}
  (2022) 008} [\href{https://arxiv.org/abs/2112.12828}{{\ttfamily
  2112.12828}}].

\bibitem{Chandrasekaran:2022cip}
V.~Chandrasekaran, R.~Longo, G.~Penington and E.~Witten, \emph{{An algebra of
  observables for de Sitter space}},
  \href{https://doi.org/10.1007/JHEP02(2023)082}{\emph{JHEP} {\bfseries 02}
  (2023) 082} [\href{https://arxiv.org/abs/2206.10780}{{\ttfamily
  2206.10780}}].

\bibitem{Chandrasekaran:2022eqq}
V.~Chandrasekaran, G.~Penington and E.~Witten, \emph{{Large N algebras and
  generalized entropy}},
  \href{https://doi.org/10.1007/JHEP04(2023)009}{\emph{JHEP} {\bfseries 04}
  (2023) 009} [\href{https://arxiv.org/abs/2209.10454}{{\ttfamily
  2209.10454}}].

\bibitem{Penington:2023dql}
G.~Penington and E.~Witten, \emph{{Algebras and States in JT Gravity}},
  \href{https://arxiv.org/abs/2301.07257}{{\ttfamily 2301.07257}}.

\bibitem{Witten:2023qsv}
E.~Witten, \emph{{Algebras, Regions, and Observers}},
  \href{https://arxiv.org/abs/2303.02837}{{\ttfamily 2303.02837}}.

\bibitem{Strohmaier:2023opz}
A.~Strohmaier and E.~Witten, \emph{{The Timelike Tube Theorem in Curved
  Spacetime}},  \href{https://arxiv.org/abs/2303.16380}{{\ttfamily
  2303.16380}}.

\bibitem{Witten:2023xze}
E.~Witten, \emph{{A background-independent algebra in quantum gravity}},
  \href{https://doi.org/10.1007/JHEP03(2024)077}{\emph{JHEP} {\bfseries 03}
  (2024) 077} [\href{https://arxiv.org/abs/2308.03663}{{\ttfamily
  2308.03663}}].

\bibitem{Jensen:2023yxy}
K.~Jensen, J.~Sorce and A.J.~Speranza, \emph{{Generalized entropy for general
  subregions in quantum gravity}},
  \href{https://doi.org/10.1007/JHEP12(2023)020}{\emph{JHEP} {\bfseries 12}
  (2023) 020} [\href{https://arxiv.org/abs/2306.01837}{{\ttfamily
  2306.01837}}].

\bibitem{Sorce:2023fdx}
J.~Sorce, \emph{{Notes on the type classification of von Neumann algebras}},
  \href{https://doi.org/10.1142/S0129055X24300024}{\emph{Rev. Math. Phys.}
  {\bfseries 36} (2024) 2430002}
  [\href{https://arxiv.org/abs/2302.01958}{{\ttfamily 2302.01958}}].

\bibitem{Xu:2024hoc}
J.~Xu, \emph{{Von Neumann Algebras in Double-Scaled SYK}},
  \href{https://arxiv.org/abs/2403.09021}{{\ttfamily 2403.09021}}.

\bibitem{Berkooz2018}
M.~Berkooz, P.~Narayan and J.~Sim{\'o}n, \emph{Chord diagrams, exact
  correlators in spin glasses and black hole bulk reconstruction},
  \href{https://doi.org/10.1007/JHEP08(2018)192}{\emph{Journal of High Energy
  Physics} {\bfseries 2018} (2018) 192}.

\bibitem{Berkooz2019}
M.~Berkooz, M.~Isachenkov, V.~Narovlansky and G.~Torrents, \emph{Towards a full
  solution of the large n double-scaled syk model},
  \href{https://doi.org/10.1007/JHEP03(2019)079}{\emph{Journal of High Energy
  Physics} {\bfseries 2019} (2019) 79}.

\bibitem{Berkooz2020SUSY}
M.~Berkooz, N.~Brukner, V.~Narovlansky and A.~Raz, \emph{The double scaled
  limit of super-symmetric syk models},
  \href{https://doi.org/10.1007/JHEP12(2020)110}{\emph{Journal of High Energy
  Physics} {\bfseries 2020} (2020) 110}.

\bibitem{Berkooz2021complex}
M.~Berkooz, V.~Narovlansky and H.~Raj, \emph{Complex sachdev-ye-kitaev model in
  the double scaling limit},
  \href{https://doi.org/10.1007/JHEP02(2021)113}{\emph{Journal of High Energy
  Physics} {\bfseries 2021} (2021) 113}.

\bibitem{Berkooz2022Quantum}
M.~Berkooz, M.~Isachenkov, P.~Narayan and V.~Narovlansky, \emph{{Quantum
  groups, non-commutative $AdS_2$, and chords in the double-scaled SYK model}},
   \href{https://arxiv.org/abs/2212.13668}{{\ttfamily 2212.13668}}.

\bibitem{Mukhametzhanov2023tcg}
B.~Mukhametzhanov, \emph{{Large p SYK from chord diagrams}},
  \href{https://arxiv.org/abs/2303.03474}{{\ttfamily 2303.03474}}.

\bibitem{Okuyama2023bch}
K.~Okuyama and K.~Suzuki, \emph{{Correlators of double scaled SYK at
  one-loop}}, \href{https://doi.org/10.1007/JHEP05(2023)117}{\emph{JHEP}
  {\bfseries 05} (2023) 117}
  [\href{https://arxiv.org/abs/2303.07552}{{\ttfamily 2303.07552}}].

\bibitem{Okuyama2023iwu}
K.~Okuyama, \emph{{High temperature expansion of double scaled SYK}},
  \href{https://arxiv.org/abs/2304.01522}{{\ttfamily 2304.01522}}.

\bibitem{Okuyama2023byh}
K.~Okuyama, \emph{{End of the world brane in double scaled SYK}},
  \href{https://arxiv.org/abs/2305.12674}{{\ttfamily 2305.12674}}.

\bibitem{Goel2023svz}
A.~Goel, V.~Narovlansky and H.~Verlinde, \emph{{Semiclassical geometry in
  double-scaled SYK}},  \href{https://arxiv.org/abs/2301.05732}{{\ttfamily
  2301.05732}}.

\bibitem{Boruch2023}
J.~Boruch, H.W.~Lin and C.~Yan, \emph{Exploring supersymmetric wormholes in n=2
  syk with chords},
  \href{https://doi.org/10.1007/JHEP12(2023)151}{\emph{Journal of High Energy
  Physics} {\bfseries 2023} (2023) 151}.

\bibitem{Verlinde:2024zrh}
H.~Verlinde and M.~Zhang, \emph{{SYK Correlators from 2D Liouville-de Sitter
  Gravity}},  \href{https://arxiv.org/abs/2402.02584}{{\ttfamily 2402.02584}}.

\bibitem{Verlinde:2024znh}
H.~Verlinde, \emph{{Double-scaled SYK, Chords and de Sitter Gravity}},
  \href{https://arxiv.org/abs/2402.00635}{{\ttfamily 2402.00635}}.

\bibitem{Narovlansky:2023lf}
V.~Narovlansky and H.~Verlinde, \emph{{Double-scaled SYK and de Sitter
  Holography}},  \href{https://arxiv.org/abs/2310.16994}{{\ttfamily
  2310.16994}}.

\bibitem{Milekhin:2023bjv}
A.~Milekhin and J.~Xu, \emph{{Revisiting Brownian SYK and its possible
  relations to de Sitter}},  \href{https://arxiv.org/abs/2312.03623}{{\ttfamily
  2312.03623}}.

\bibitem{Lin2022}
H.W.~Lin, \emph{The bulk hilbert space of double scaled syk},
  \href{https://doi.org/10.1007/JHEP11(2022)060}{\emph{Journal of High Energy
  Physics} {\bfseries 2022} (2022) 60}.

\bibitem{Susskind2021omt}
L.~Susskind, \emph{{De Sitter Holography: Fluctuations, Anomalous Symmetry, and
  Wormholes}}, \href{https://doi.org/10.3390/universe7120464}{\emph{Universe}
  {\bfseries 7} (2021) 464} [\href{https://arxiv.org/abs/2106.03964}{{\ttfamily
  2106.03964}}].

\bibitem{Susskind2021dfc}
L.~Susskind, \emph{{Black Holes Hint Towards De Sitter-Matrix Theory}},
  \href{https://arxiv.org/abs/2109.01322}{{\ttfamily 2109.01322}}.

\bibitem{Susskind2021esx}
L.~Susskind, \emph{{Entanglement and Chaos in De Sitter Space Holography: An
  SYK Example}}, \href{https://doi.org/10.22128/jhap.2021.455.1005}{\emph{JHAP}
  {\bfseries 1} (2021) 1} [\href{https://arxiv.org/abs/2109.14104}{{\ttfamily
  2109.14104}}].

\bibitem{Susskind2022fop}
E.~Shaghoulian and L.~Susskind, \emph{{Entanglement in De Sitter space}},
  \href{https://doi.org/10.1007/JHEP08(2022)198}{\emph{JHEP} {\bfseries 08}
  (2022) 198} [\href{https://arxiv.org/abs/2201.03603}{{\ttfamily
  2201.03603}}].

\bibitem{Susskind2022dfz}
L.~Susskind, \emph{{Scrambling in Double-Scaled SYK and De Sitter Space}},
  \href{https://arxiv.org/abs/2205.00315}{{\ttfamily 2205.00315}}.

\bibitem{SusskindLin2022nss}
H.~Lin and L.~Susskind, \emph{{Infinite Temperature's Not So Hot}},
  \href{https://arxiv.org/abs/2206.01083}{{\ttfamily 2206.01083}}.

\bibitem{Susskind2022bia}
L.~Susskind, \emph{{De Sitter Space, Double-Scaled SYK, and the Separation of
  Scales in the Semiclassical Limit}},
  \href{https://arxiv.org/abs/2209.09999}{{\ttfamily 2209.09999}}.

\bibitem{Susskind2023hnj}
L.~Susskind, \emph{{De Sitter Space has no Chords. Almost Everything is
  Confined}},  \href{https://arxiv.org/abs/2303.00792}{{\ttfamily 2303.00792}}.

\bibitem{Susskind2023rxm}
L.~Susskind, \emph{{A Paradox and its Resolution Illustrate Principles of de
  Sitter Holography}},  \href{https://arxiv.org/abs/2304.00589}{{\ttfamily
  2304.00589}}.

\bibitem{Jackiw1984je}
R.~Jackiw, \emph{{Lower Dimensional Gravity}},
  \href{https://doi.org/10.1016/0550-3213(85)90448-1}{\emph{Nucl. Phys. B}
  {\bfseries 252} (1985) 343}.

\bibitem{Teitelboim1983ux}
C.~Teitelboim, \emph{{Gravitation and Hamiltonian Structure in Two Space-Time
  Dimensions}}, \href{https://doi.org/10.1016/0370-2693(83)90012-6}{\emph{Phys.
  Lett. B} {\bfseries 126} (1983) 41}.

\bibitem{Polchinski2015}
A.~Almheiri and J.~Polchinski, \emph{Models of ads2 backreaction and
  holography}, \href{https://doi.org/10.1007/JHEP11(2015)014}{\emph{Journal of
  High Energy Physics} {\bfseries 2015} (2015) 14}.

\bibitem{jensen2016}
K.~Jensen, \emph{Chaos in ${\mathrm{ads}}_{2}$ holography},
  \href{https://doi.org/10.1103/PhysRevLett.117.111601}{\emph{Phys. Rev. Lett.}
  {\bfseries 117} (2016) 111601}.

\bibitem{yang2016}
J.~Maldacena, D.~Stanford and Z.~Yang, \emph{{Conformal symmetry and its
  breaking in two-dimensional nearly anti-de Sitter space}},
  \href{https://doi.org/10.1093/ptep/ptw124}{\emph{Progress of Theoretical and
  Experimental Physics} {\bfseries 2016} (2016) }
  [\href{https://arxiv.org/abs/https://academic.oup.com/ptep/article-pdf/2016/12/12C104/9620935/ptw124.pdf}{{\ttfamily
  https://academic.oup.com/ptep/article-pdf/2016/12/12C104/9620935/ptw124.pdf}}].

\bibitem{Mertens2022}
T.G.~Mertens and G.J.~Turiaci, \emph{{Solvable models of quantum black holes: a
  review on Jackiw\textendash{}Teitelboim gravity}},
  \href{https://doi.org/10.1007/s41114-023-00046-1}{\emph{Living Rev. Rel.}
  {\bfseries 26} (2023) 4} [\href{https://arxiv.org/abs/2210.10846}{{\ttfamily
  2210.10846}}].

\bibitem{Saad:2019lba}
P.~Saad, S.H.~Shenker and D.~Stanford, \emph{{JT gravity as a matrix
  integral}},  \href{https://arxiv.org/abs/1903.11115}{{\ttfamily 1903.11115}}.

\bibitem{Jafferis:2019wkd}
D.L.~Jafferis and D.K.~Kolchmeyer, \emph{{Entanglement Entropy in
  Jackiw-Teitelboim Gravity}},
  \href{https://arxiv.org/abs/1911.10663}{{\ttfamily 1911.10663}}.

\bibitem{Harlow:2018tqv}
D.~Harlow and D.~Jafferis, \emph{{The Factorization Problem in
  Jackiw-Teitelboim Gravity}},
  \href{https://doi.org/10.1007/JHEP02(2020)177}{\emph{JHEP} {\bfseries 02}
  (2020) 177} [\href{https://arxiv.org/abs/1804.01081}{{\ttfamily
  1804.01081}}].

\bibitem{Takayanagi2006}
S.~Ryu and T.~Takayanagi, \emph{Holographic derivation of entanglement entropy
  from the anti--de sitter space/conformal field theory correspondence},
  \href{https://doi.org/10.1103/PhysRevLett.96.181602}{\emph{Phys. Rev. Lett.}
  {\bfseries 96} (2006) 181602}.

\bibitem{kitaev2015}
A.Kitaev, \emph{A simple model of quantum holography},  2015.

\bibitem{sachdev1993}
S.~Sachdev and J.~Ye, \emph{Gapless spin-fluid ground state in a random quantum
  heisenberg magnet},
  \href{https://doi.org/10.1103/PhysRevLett.70.3339}{\emph{Phys. Rev. Lett.}
  {\bfseries 70} (1993) 3339}.

\bibitem{maldacena2016remarks}
J.~Maldacena and D.~Stanford, \emph{Remarks on the sachdev-ye-kitaev model},
  \href{https://doi.org/10.1103/PhysRevD.94.106002}{\emph{Phys. Rev. D}
  {\bfseries 94} (2016) 106002}.

\bibitem{Polchinski2016}
J.~Polchinski and V.~Rosenhaus, \emph{The spectrum in the sachdev-ye-kitaev
  model}, \href{https://doi.org/10.1007/JHEP04(2016)001}{\emph{Journal of High
  Energy Physics} {\bfseries 2016} (2016) 1}.

\bibitem{altland2016}
D.~Bagrets, A.~Altland and A.~Kamenev, \emph{Sachdev–ye–kitaev model as
  liouville quantum mechanics},
  \href{https://doi.org/https://doi.org/10.1016/j.nuclphysb.2016.08.002}{\emph{Nuclear
  Physics B} {\bfseries 911} (2016) 191}.

\bibitem{altland2017}
D.~Bagrets, A.~Altland and A.~Kamenev, \emph{Power-law out of time order
  correlation functions in the syk model},
  \href{https://doi.org/https://doi.org/10.1016/j.nuclphysb.2017.06.012}{\emph{Nuclear
  Physics B} {\bfseries 921} (2017) 727}.

\bibitem{mehta2004random}
M.~Mehta, \emph{Random Matrices}, ISSN, Elsevier Science (2004).

\bibitem{Garcia-Garcia:2018fns}
A.M.~Garc\'\i{}a-Garc\'\i{}a, Y.~Jia and J.J.M.~Verbaarschot, \emph{{Exact
  moments of the Sachdev-Ye-Kitaev model up to order $1/N^2$}},
  \href{https://doi.org/10.1007/JHEP04(2018)146}{\emph{JHEP} {\bfseries 04}
  (2018) 146} [\href{https://arxiv.org/abs/1801.02696}{{\ttfamily
  1801.02696}}].

\bibitem{Lin2023}
H.W.~Lin and D.~Stanford, \emph{{A symmetry algebra in double-scaled SYK}},
  \href{https://doi.org/10.21468/SciPostPhys.15.6.234}{\emph{SciPost Phys.}
  {\bfseries 15} (2023) 234}.

\bibitem{Parker2019}
D.E.~Parker, X.~Cao, A.~Avdoshkin, T.~Scaffidi and E.~Altman, \emph{A universal
  operator growth hypothesis},
  \href{https://doi.org/10.1103/PhysRevX.9.041017}{\emph{Phys. Rev. X}
  {\bfseries 9} (2019) 041017}.

\bibitem{Bhattacharyya2023}
A.~Bhattacharyya, S.S.~Haque, G.~Jafari, J.~Murugan and D.~Rapotu, \emph{Krylov
  complexity and spectral form factor for noisy random matrix models},
  \href{https://doi.org/10.1007/JHEP10(2023)157}{\emph{Journal of High Energy
  Physics} {\bfseries 2023} (2023) 157}.

\bibitem{Erdmenger2023}
J.~Erdmenger, S.-K.~Jian and Z.-Y.~Xian, \emph{Universal chaotic dynamics from
  krylov space}, \href{https://doi.org/10.1007/jhep08(2023)176}{\emph{Journal
  of High Energy Physics} {\bfseries 2023} (2023) }.

\bibitem{Kar2022}
A.~Kar, L.~Lamprou, M.~Rozali and J.~Sully, \emph{Random matrix theory for
  complexity growth and black hole interiors},
  \href{https://doi.org/10.1007/JHEP01(2022)016}{\emph{Journal of High Energy
  Physics} {\bfseries 2022} (2022) 16}.

\bibitem{Rabinovici2021}
E.~Rabinovici, A.~S{\'a}nchez-Garrido, R.~Shir and J.~Sonner, \emph{Operator
  complexity: a journey to the edge of krylov space},
  \href{https://doi.org/10.1007/JHEP06(2021)062}{\emph{Journal of High Energy
  Physics} {\bfseries 2021} (2021) 62}.

\bibitem{Jian2021}
S.-K.~Jian, B.~Swingle and Z.-Y.~Xian, \emph{Complexity growth of operators in
  the syk model and in jt gravity},
  \href{https://doi.org/10.1007/JHEP03(2021)014}{\emph{Journal of High Energy
  Physics} {\bfseries 2021} (2021) 14}.

\bibitem{Rabinovici20221}
E.~Rabinovici, A.~S{\'a}nchez-Garrido, R.~Shir and J.~Sonner, \emph{Krylov
  localization and suppression of complexity},
  \href{https://doi.org/10.1007/JHEP03(2022)211}{\emph{Journal of High Energy
  Physics} {\bfseries 2022} (2022) 211}.

\bibitem{Rabinovici20222}
E.~Rabinovici, A.~S{\'a}nchez-Garrido, R.~Shir and J.~Sonner, \emph{Krylov
  complexity from integrability to chaos},
  \href{https://doi.org/10.1007/JHEP07(2022)151}{\emph{Journal of High Energy
  Physics} {\bfseries 2022} (2022) 151}.

\bibitem{Liu2023}
C.~Liu, H.~Tang and H.~Zhai, \emph{Krylov complexity in open quantum systems},
  \href{https://doi.org/10.1103/PhysRevResearch.5.033085}{\emph{Phys. Rev.
  Res.} {\bfseries 5} (2023) 033085}.

\bibitem{Tang:2023ocr}
H.~Tang, \emph{{Operator Krylov complexity in random matrix theory}},
  \href{https://arxiv.org/abs/2312.17416}{{\ttfamily 2312.17416}}.

\end{thebibliography}\endgroup
\end{document}